\documentclass[aps,prl,floatfix,superscriptaddress,10pt,twocolumn]{revtex4}
\usepackage{color,amsmath,amssymb,graphicx,latexsym,subfigure}
\usepackage{threeparttable,multirow,txfonts,makecell,tabularx}
\DeclareMathAlphabet{\mathcal}{OMS}{cmsy}{m}{n}
\usepackage{appendix}
\usepackage{hyperref}
\newcommand{\nc}{\newcommand}
\nc{\beqa}{\begin{eqnarray}}  
\nc{\eeqa}{\end{eqnarray}}

\begin{document}

\title{Solving the domain wall problem with first-order phase transition}

\author{Yang Li }
\affiliation{School of Physical Sciences, University of Chinese Academy of Sciences, No. 19A Yuquan Road, Beijing 100049, China}
\affiliation{
CAS Key Laboratory of Theoretical Physics, Institute of Theoretical Physics, Chinese Academy of Sciences,
P.O. Box 2735, Beijing 100190, China}

\author{Ligong Bian }\email{lgbycl@cqu.edu.cn}
\affiliation{Department of Physics and Chongqing Key Laboratory for Strongly Coupled Physics, Chongqing University, Chongqing 401331, P. R. China}
\affiliation{ Center for High Energy Physics, Peking University, Beijing 100871, China}

\author{Yongtao Jia }
\affiliation{Department of Physics and Chongqing Key Laboratory for Strongly Coupled Physics, Chongqing University, Chongqing 401331, P. R. China}

\begin{abstract}

Domain wall networks are two-dimensional topological defects generally predicted in many beyond standard model physics. In this Letter, we propose to solve the domain wall problem with the first-order phase transition. We numerically study the phase transition dynamics, and for the first time show that the domain walls reached scaling regime can be diluted through the interaction with vacuum bubbles during the first-order phase transition. We find that the amplitude of the gravitational waves produced by the second-stage first-order phase transition is several orders higher than that from the domain walls evolution in the scaling regime. The scale of the first-order phase transition that dilute the domain walls can be probed through gravitational waves detection.

\end{abstract}

\maketitle

\noindent{\it \bfseries  Introduction.}
Domain walls (DWs) are two-dimensional surface-like topological defects forming after the spontaneous breakdown of a 
discrete symmetry in the early Universe~\cite{kibble1976topology,zurek1985cosmological,Zeldovich:1974uw}. DWs are generally predicted in many particle physics models, such as axion models~\cite{Sikivie:1982qv,Vilenkin:1982ks,Hiramatsu:2010yn,Hiramatsu:2012sc,Kawasaki:2014sqa}, supersymmetric models~\cite{Abel:1995wk,Takahashi:2008mu,Dine:2010eb,Hamaguchi:2011nm,Kadota:2015dza}, thermal inflation~\cite{Moroi:2011be}, etc. 
 If the domain wall networks are stable they would eventually overclose the universe and therefore 
conflict with the present 
cosmological observations, since the energy density of DWs decreases slower than that of matters and radiations, this is the well-known domain wall problem~ \cite{Zeldovich:1974uw}.
The conventional way to resolve this problem is to introduce a biased potential to lift the degenerate minima so that the walls will annihilate at sufficiently early
times~\cite{Coulson:1995nv,Gelmini:1988sf,Vilenkin:1981zs,Larsson:1996sp,sikivie1982axions} and remains the stochastic gravitational wave background (SGWB)\cite{Gleiser:1998na,Hiramatsu:2010yz,Kawasaki:2011vv,Hiramatsu:2013qaa}. More recently, the PPTA and NANOGrav datasets are used to place constraints on the SGWB produced emitted by DW networks~\cite{Bian:2022qbh,Ferreira:2022zzo}.

In this Letter, we propose to evade the domain wall problem with the following first-order phase transition (PT) process after the DWs are formed. 
Recent development of gravitational wave (GW) astronomy provides a distinctive way to probe particle physics predicting first-order PT in the early Universe~\cite{Caprini:2015zlo,Bian:2021ini,Caprini:2019egz,Cai:2017cbj,Caldwell:2022qsj}. Recent constraints on the low-scale phase transitions are placed with the dataset of Parkes Pulsar Timing Array (PPTA) and NANOGrav collabrations~\cite{Bian:2020urb,Xue:2021gyq,NANOGrav:2021flc}. 
The first constraints on the PeV-EeV scale phase transitions are reached in Ref.~\cite{Romero:2021kby} by utilizing the data from
the third observing runs of LIGO-Virgo.

In particular, we will focus on two-step PTs where the DW defects are created in the first step and the first-order PT happens in the second step. Both the first-step second-order PT together with DWs evolution and the second-step first-order PT are expected to generate GWs.
When DWs are formed after the first-step second order PT, they would evolve toward the “scaling regime” where the typical length scales of the DWs network
( the curvature radius and the distance between neighboring DWs) become comparable to the Hubble radius ($H^{-1}$). In this regime, DWs would frequently interact with each other, changing their configuration or collapsing
into the closed walls to maintain the scaling property, and yields GW generation~\cite{Hiramatsu:2010yz,Kawasaki:2011vv,Hiramatsu:2013qaa}. The scenario where the first-order PT leads to the breakdown of the electroweak symmetry is well motivated for dark matter and baryon asymmetry of the Universe\cite{McDonald:1993ex,Burgess:2000yq,Espinosa:2007qk,Profumo:2007wc,Barger:2007im,Espinosa:2008kw,Espinosa:2011ax,Espinosa:2011eu,Cline:2012hg,Profumo:2014opa,Feng:2014vea,
Curtin:2014jma,Craig:2014lda,Huang:2016cjm,Vaskonen:2016yiu,Curtin:2016urg,Kurup:2017dzf,Buttazzo:2018qqp,Caprini:2019egz,Alanne:2019bsm,Costantini:2020stv,AlAli:2021let}.  

We numerically study the formation and decay of the DWs followed by a first-order PT, 
and find that vacuum bubbles would expand and collide with each other and DWs. We characterize the DWs behavior with scaling parameters by including the interplay between vacuum bubbles and DWs for the first time. The scaling parameter of DWs indicates that the DWs decay is a natural result of the collision between DW networks and vacuum bubbles during the second stage. After vacuum bubbles are generated,  vacuum bubbles would expand and collide with each other and domain walls to generate GWs. We also find that the first-order PTs dominate the amplitude and shape of the GW spectrum.

\noindent{\it \bfseries  The model.}
For simplicity and without loss of universality, we consider the domain walls being formed after spontaneous symmetry breaking of a $Z_2$ symmetry, and 
the second step is a first-order phase transition to restore the $Z_2$ symmetry. With the high-temperature approximation, the thermal potential keeping only the terms up to $\propto T^2$ is given by,
\begin{eqnarray}
V(S,h)&=&- \frac{1}{2}(a v^2 - c_h T^2) h^2 +\frac{\lambda_h}{4} h^4   \nonumber\\
&&-  \frac{1}{2} (b v^2 - c_S T^2)s^2+\frac{\lambda_S}{4}  s^4+\frac{\lambda_{Sh}}{2} h^2s^2,
\end{eqnarray}
with $\lambda_{Sh,S,h} >0$, and $v$ being a scale parameter. For the case where the second step is a spontaneous EW symmetry-breaking process, one has $a v^2=\mu_h^2,b v^2=\mu_S^2$, and the thermal effects are captured by $c_h$ and $c_s$~\cite{Espinosa:2011ax}:
\begin{equation}
 c_h = \frac{2 m_W^2 + m_Z^2 + m_h^2 + 2 m_t^2}{4 v^2}
 + \frac{\lambda_{Sh}}{12}, \quad c_S= \frac{ 4 \lambda_{Sh} + 3 \lambda_S}{12},
\end{equation}
With the thermal potential, all symmetries are restored at high temperatures. 
As the Universe cools down, the discrete $Z_2$ symmetry firstly breakdown spontaneously with the singlet scalar $S$ developing a non-vanishing vacuum expectation value 
$\langle S \rangle = \pm v_s= \pm\mu_S/\sqrt{\lambda_S}$, with the space regions equivalently fails into $+$ and $-$ domains formed inside a Hubble patch
separated by domain walls at the boundaries: $S_\text{DW}(z) = v_s \,\text{tanh}(\mu_S z/\sqrt{2})$.
 Later, the DWs would enter into the scaling regime or not depending on if the time separation between the first-step and the second-step PTs is long enough. Then, the first-step PT occurs equivalently
either inside the $+$ or $-$ domains, with the vacua of $(\langle h \rangle,\langle S \rangle)$ transiting as $(0, \pm v_s) \rightarrow (v_h=\mu_h/\sqrt{\lambda_h},0)$. Since we intend to study the dynamics from the formation of domains walls and the effects of the first order PT on the evolution of DWs network, we do not consider bias term to appear as in Ref.~\cite{McDonald:1995hp,Espinosa:2011eu}. For the study of the seeded vacuum
decay by domain walls, we refer to Ref.~\cite{Blasi:2022woz,Blasi:2023rqi}, where the formed domain walls are required not to collapse before the second-step PT
has been completed.
We note that to obtain the desired phase transition the following two requirements are necessary~\cite{Bian:2018mkl}: 1) at zero temperature, the local vacuum at $(h,S)=(0,\pm v_S)$ should be higher the global one at $(h,S)=
(v_h,0)$, which yields $\lambda_S\lambda_h>2\mu_S^4/(2\mu_h^4)$ with $\mu_h^2, \mu_S^2 >0$; and 2) the first step second-order phase transition occurs earlier than the second-step first-order phase transition: $\mu_h^2/c_h<\mu_S^2/c_S$.

\noindent{\it \bfseries  The simulation framework.}
At high temperatures before the phase transition occurs in the radiation-dominant universe,
we consider the scalar field $S$ evolving from their initial thermal configuration until the critical temperature when the second-order PT occurs, and the amplitude and momentum of the field $h$ initially have vanishing values. As the temperature future cools down, the quantum tunneling process realizes through vacuum bubbles nucleation in the second-step first-order PT. During the numerical simulation of the
classical scalar fields, the transverse-traceless(TT) part of the anisotropic energy-momentum tensor 
($\Pi^{TT}_{ij} = (\partial_i s \partial_j s+\partial_i h \partial_j h)$) would induce the spatial metric perturbations $h_{ij}$ and contribute to GWs energy density $\rho_{gw}=\langle\dot{h}_{ij}(x,t)\dot{h}_{ij}(x,t)\rangle_V$ with $\langle...\rangle_V$ denoting spatial average over the whole simulation volume.  To eliminate the influence of the initial conditions (especially different initial temperatures) on the GWs power spectrum, we start to evolve the spatial metric perturbations ($h_{ij}$) from the critical temperature of the second-orderPT (see {\it Supplemental material} for details). 

 We intend to study the case where DWs reach the scaling regime before the first-order PT, and the parameters in the thermal potential are chosen as follows: $\lambda_h=0.129$, $\lambda_{Sh}=0.66$, $\lambda_S=1$, $a=0.781$, $c_h=0.4534$, $b=1.104$, $c_S=0.235$. Therefore, the critical temperature of second-order PT is about $T_{\rm c}=2.17v$.
 For the convenience of numerical simulation, all scalar fields $(S,h)$ are rescaled with the scale parameter $v$, and the grid size $dx$ (and time-step) are rescaled by $w_*=a_iH_i$ with the $a_i$ and $H_i$ being initial scale factor and Hubble parameter. It is convenient to take the rescaled conformal time $\tilde{\eta}(=\eta/\eta_i)$ with the initial conformal time being $\eta_i=1/w_*$ in our simulations. 
 We fix $\tilde{\eta}=1$ to be the initial time at which $T=12T_{\rm c}$, so the second-order PT happens at $\tilde{\eta}=12$. We set $v=6\times10^{16}{\rm GeV}$ to ensure the simulation box can capture large enough Hubble volume and have enough time to evolve the DWs network.

We use the second-order leap-frog algorithm adopted by ${\mathcal CosmoLattice}$~\cite{Figueroa:2021yhd,Figueroa:2020rrl} to evolve the equations of motion in a simulation box of comoving side-length $L=190/(a_{i}H_{i})$ and $N=1024$ points per side. So, the dimensionless comoving lattice spacing is about $\delta \Tilde{x}=0.19$, and the dimensionless time-step is chosen as $\delta \tilde{\eta}=0.005$. 
As the temperature decreases to about $T=0.5v$ ( $\tilde{\eta}=52.38$) when the domain walls have entered the scaling regime, the quantum tunneling process of the first-order PT occurs, the bubble profile functions of $s$ and $h$ fields have the following form
\begin{gather}
s(r)_{\pm}=\pm 0.7v \pm \frac{s_v}{2}\left( 1+{\rm tanh}\left[ \frac{r-(R_{\rm in}+0.5L_{\rm w})}{L_{\rm w}} \right] \right),\\  \nonumber
\\ 
h(r)= \frac{h_v}{2}\left( 1-{\rm tanh}\left[ \frac{r-(R_{\rm in}+0.5L_{\rm w})}{L_{\rm w}} \right] \right),
\end{gather}
where $s_v=0.32v$, $h_v=1.1v$, $r \in [0,25]\delta x_{\rm phy}$, $R_{\rm in}=10\delta x_{\rm phy}$, $L_{\rm w}=12\delta x_{\rm phy}$, $\delta x_{\rm phy}$ is the physical lattice spacing when bubble is generated. The PT strength (the latent heat normalized by the radiation energy) can be estimated to be $\alpha=(V(0,\langle s\rangle)-V(\langle h\rangle,0))/\rho_r\approx0.276$. We randomly generate bubbles far from the DWs. For our simulation, the vacuum bubbles can be considered to generate simultaneously with the bubble nucleation rate $\beta\approx 8\pi (N_b/\mathcal{V})^{1/3}$ with the $N_b$ and $\mathcal{V}$ being bubbles number and the volume of the simulation box~\cite{Hindmarsh:2020hop}.
We place the different number of bubbles in the simulation box to study the effect of the number of bubbles (denote as $N_{\rm b}$) on the simulation, i.e. $N_{\rm b}$=512 and $N_{\rm b}$=64 in the physical volume of $V=48H_{PT}^{-3}$ which correspond to the inverse PT durations being $\beta/H_{PT}=55.3$ and $27.7$ . For concreteness, we use $s(r)_+$ to generate bubbles for the field $s$ in the spherical region where the average value of $s$ is greater than 0.9$v$, and use $s(r)_-$ to generate bubbles for the field $s$ in the spherical region where the average value of $s$ is smaller than $-0.9v$. In each spherical region that the $s$ field generates bubbles with $s(r)_+$ or $s(r)_-$, we use the $h(r)$ profile function to generate bubbles for the field $h$. 
 We evolve the equations of motion until the final time $\tilde{\eta}=150$. At this time, the simulation box contains about two Hubble volumes to reduce finite volume effects, and the thickness of the domain wall is about twice the physical lattice spacing to obtain sufficient resolution.

\begin{figure}[htb]
\includegraphics[width=.4\textwidth]{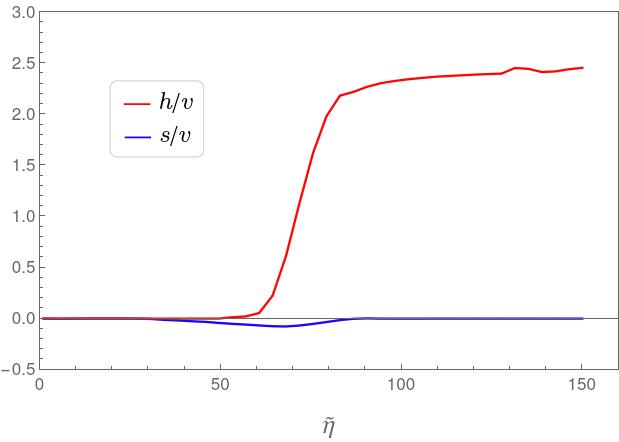} 
\hspace{16mm}
  \caption{  Evolutions of the average amplitudes of field $s$ and $h$ for the case of $N_{\rm b}=512$.  }
  \vspace{0.1cm}
  \label{fig:hsNb512}
\end{figure}

\begin{figure}[!htp]
\includegraphics[width=.2\textwidth]{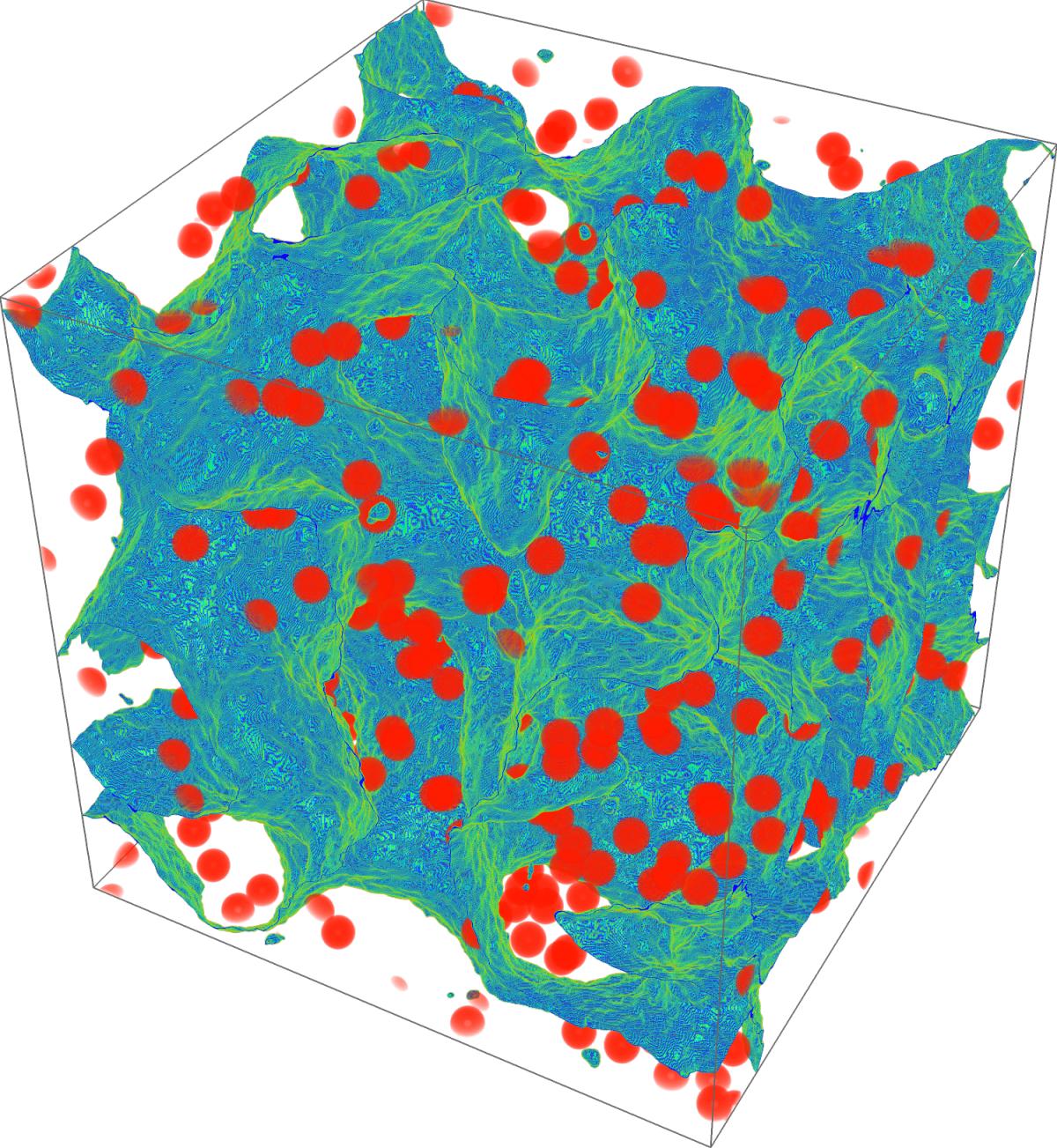} 
\includegraphics[width=.2\textwidth]{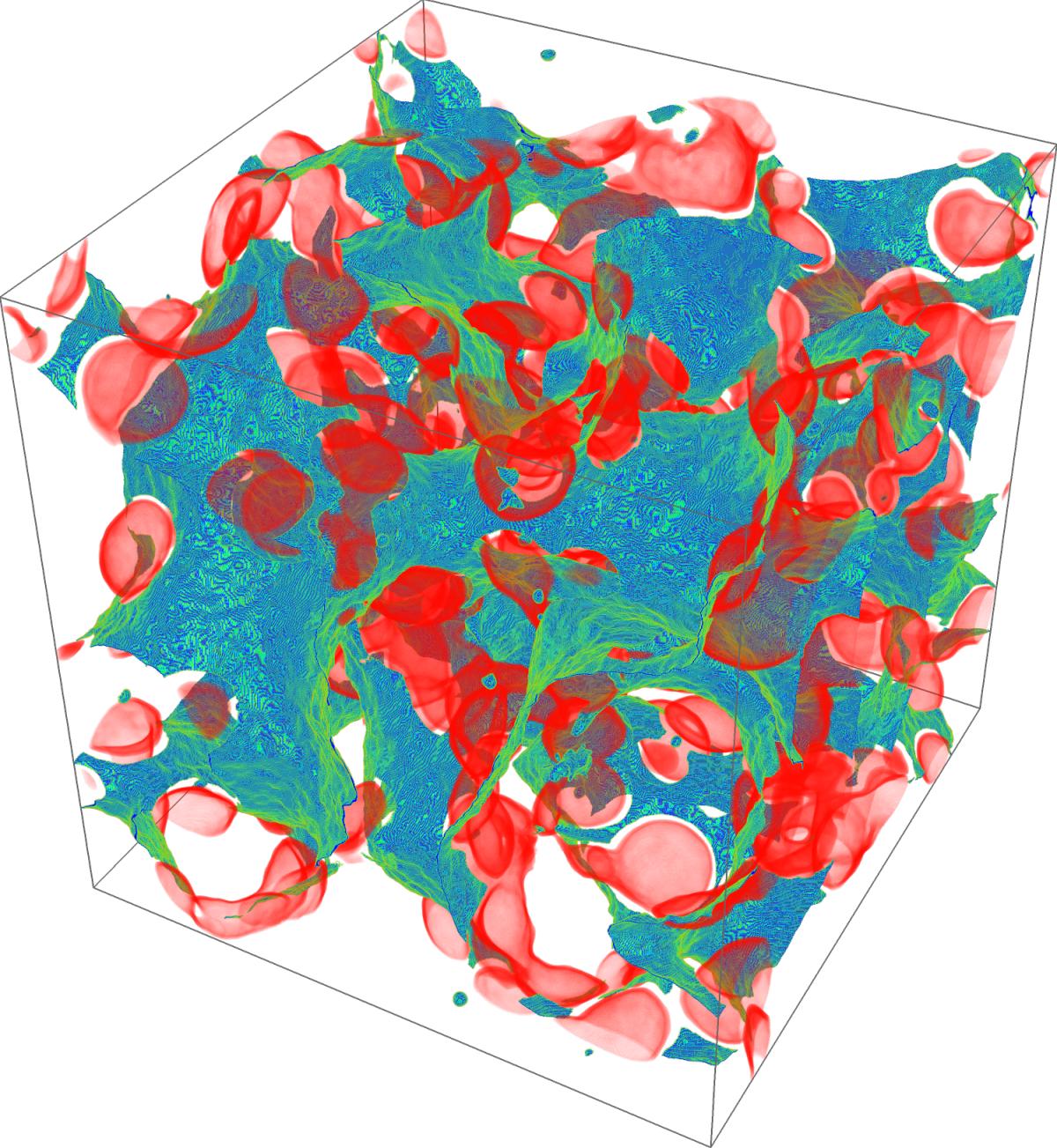}
\includegraphics[width=.2\textwidth]{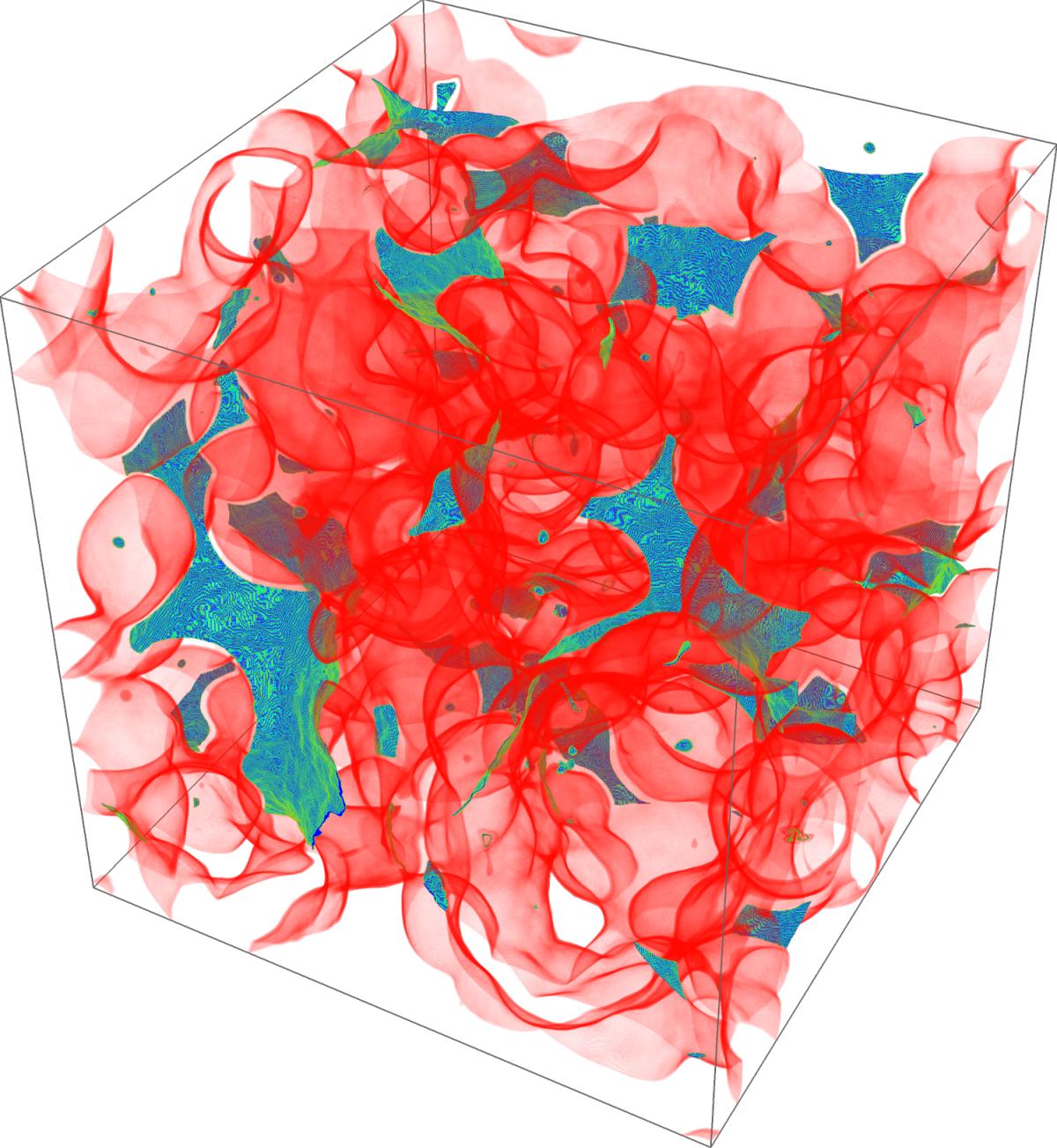}
\includegraphics[width=.2\textwidth]{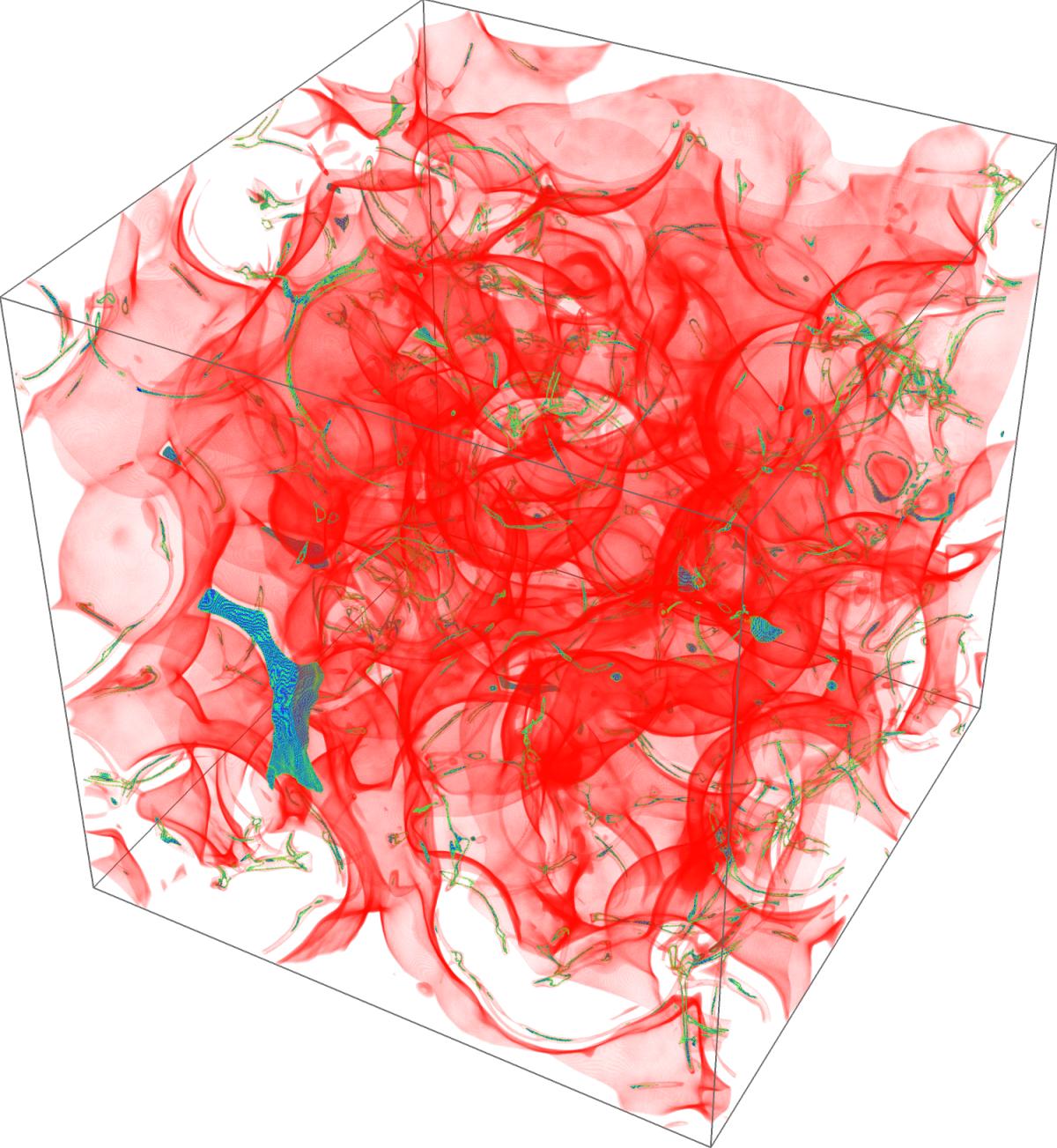}
  \caption{3D snapshots of the DWs network and the field value of $h$ at different times of $\tilde{\eta}$=52.39, 63.81, 69.52, and 75.23 from top-left to bottom-right plots. The blue region indicates where the DWs exist, and the red region indicates where the bubble of the field $h$ exists.  }
  \label{fig:3Ds}
\end{figure}

\noindent{\it \bfseries  Results.} For illustration, in Fig.~\ref{fig:hsNb512}, we present the averaged field values of $h$ and $s$ for the case of  $N_{\rm b}=512$, where we can observe that, as the universe cools down, the average amplitude of the fields $s$ and $h$ changes. The averaged amplitude of $h$ increase when the first-order PT occurs and stay stable when the phase transition is completed, and the averaged amplitude of $s$ drops below zero when the first-order PT start and back to zero when the first-order PT finished associated with the whole volume are full of vacuum: $(\langle h\rangle,\langle s\rangle)=({h_v,0})$.
The three-dimensional distribution of the DWs of the field $s$ and the bubbles of the field $h$ at different times are shown in Fig.~\ref{fig:3Ds}, which reflect 
 the physical picture of the PT dynamics. The top-left graph shows that bubbles are randomly generated away from the domain wall. The remaining three graphs show that as bubbles expand, collisions between bubbles and domain walls, as well as collisions between different bubbles cause the domain walls to decay.

\begin{figure}[!htp]
\includegraphics[width=.4\textwidth]{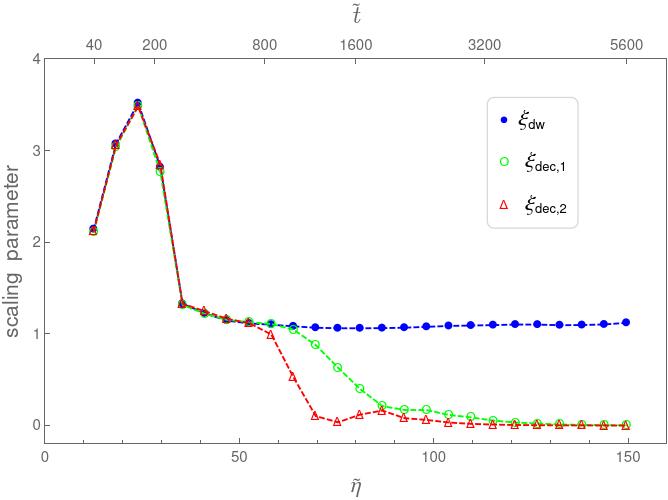} 
\hspace{16mm}
  \caption{Evolution of scaling parameters with rescaled conformal time $\tilde{\eta}$ and dimensionless cosmic time $\tilde{t}$ ($\tilde{t}=w_*/(2H)$) for three cases. The blue solid dot, the green hollow, and the red triangle points represent the scaling parameter of the DWs in the scenario of  $N_{\rm b}=0$, $N_{\rm b}=64$, and $N_{\rm b}=512$.}
  \vspace{0.1cm}
  \label{fig:scp}
\end{figure}

To reflect the impact of the number of bubbles on the domain walls' evolutions, 
we quantitatively describe the decay process of the DWs by measuring the scaling parameters
in three cases($N_{\rm b}=0, N_{\rm b}=64$ and $N_{\rm b}=512$), see Fig.~\ref{fig:scp}. For the case without considering the first-order phase transition, we obtain the scaling parameter $\xi_{\rm dw}\approx 1$, which is in agreement with previous studies of domain wall dynamics~\cite{Hiramatsu:2013qaa} except that our simulation time is much longer and the scaling parameter is more stable, see {\it Supplemental material} for details. The scaling parameter drops as the first-order PT process proceeds, and the faster PT case with more vacuum bubbles would yield an earlier decrease of the scaling parameter. 
We refer to the moment when the scaling parameter drops to half of its value in the scaling regime as the decay time of the DW. So, as can be seen from the figure. Combining the three-dimensional shape of the potential and the profile function of the bubbles of the fields $s$ and $h$, it can be inferred that due to the expansion of the bubbles, the domain wall located in the false vacuum feels the pressure difference with the true vacuum when it collides with the bubble. This pressure difference causes the false vacuum at the domain wall to be pushed to a true vacuum, which causes the decay of the domain wall. As the first-order PT complete, the scaling parameters $\xi_{\rm dec,1}$ and $\xi_{\rm dec,2}$ gradually reach zero.

\begin{figure}[htb]
\includegraphics[width=.4\textwidth]{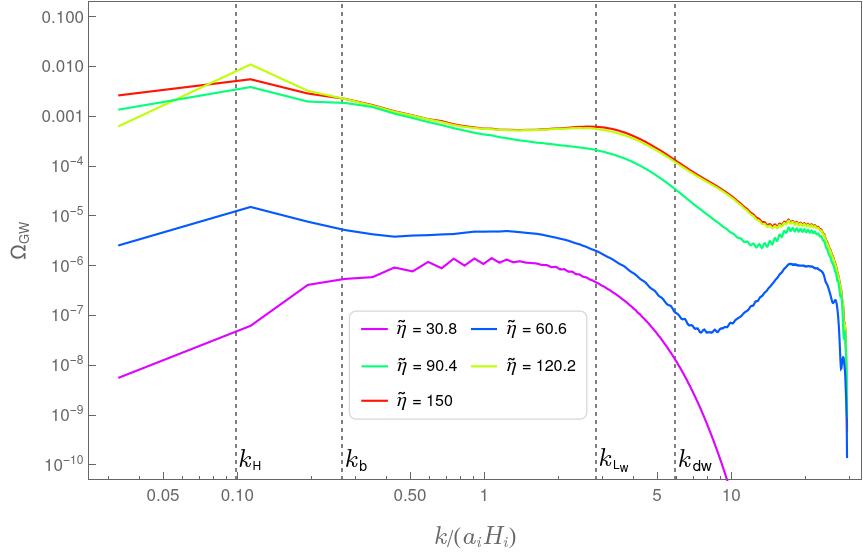} 
\hspace{16mm}
  \caption{GW power spectrum at different times in a two-step phase transition scenario with $N_{\rm b}=512$. The four vertical black dashed lines represent the characteristic scales corresponding to the Hubble radius when the domain wall decays($k_{\rm H}$), the average distance when the bubbles collide with each other ($k_b$), the thickness when the bubbles are generated($k_{L_w}$), and the thickness when the domain wall just enters the scaling regime($k_{\rm dw}$). }
  \vspace{0.1cm}
  \label{fig:gwNb512}
\end{figure}

We show the GW energy density power spectrum of the case of $N_{\rm b}=512$ in Fig.~\ref{fig:gwNb512}. The magnitude of the GWspectrum produced by domain walls at the time of $\tilde{\eta}=30.8$ is found to be around $10^{-6}$, the amplitude of the gravitational waves spectrum grows by almost four orders with first-order PT process continues and stop grows as the first-order PT complete. 
 In the figure, the four characteristic scales are given: 
$k_{\rm H}=(2\pi H_{\rm dec}) \times a(\tilde{\eta}_{\rm dec}) /(a_iH_i), \quad k_{\rm b}=2\pi/(V/N_{\rm b})^{1/3} /(a_iH_i), \quad k_{{\rm L}_{\rm w}}=2\pi/L_{\rm w} \times a(\tilde{\eta}_{\rm gen}) /(a_iH_i), \quad k_{\rm dw}=(2\pi/\delta_{\rm dw}) \times a(\tilde{\eta}_{\rm sc}) /(a_iH_i)$, which corresponds to the Hubble parameter at the time of the decay of DWs ($H_{\rm dec}$), the DWs decay time($\tilde{\eta}_{\rm dec}$=63.81), the mean bubble separation is $R_\star=(V/N_{\rm b})^{1/3}$ ($V$ is the comoving volume of the simulation box, $N_{\rm b}$ is the number of vacuum bubbles) , the wall thickness $L_{\rm w}$ when the bubbles are generated ($\tilde{\eta}_{\rm gen}$ is the time when bubbles are generated), the physical thickness of the domain wall when just entering the scaling regime ($\delta_{\rm dw}$, where $\tilde{\eta}_{\rm sc}$ is the time when domain wall just entering the scaling regime). 

\begin{figure}[htb]
\includegraphics[width=.4\textwidth]{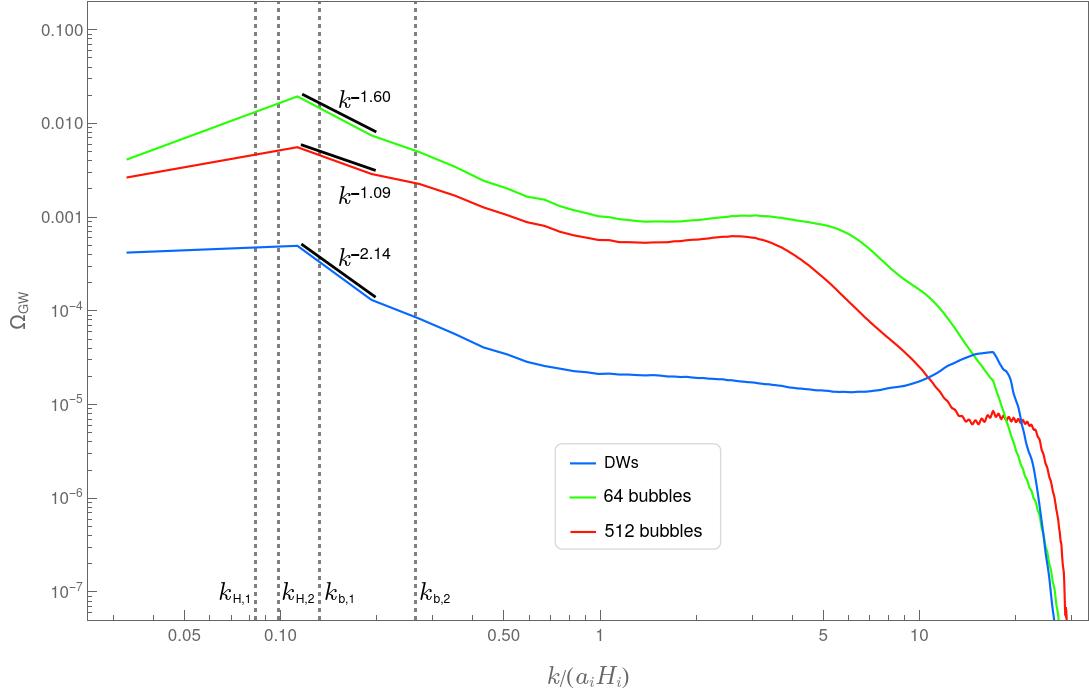} 
\hspace{16mm}
  \caption{GW power spectrum measured at the final time in three cases. The blue solid line represents the GW power spectrum measured in the scenario where only the second-order PT occurs, and the green and red solid lines are the GW power spectrum measured in the two-step PT scenario with vacuum bubble numbers $N_{\rm b}=64$ and $N_{\rm b}=512$.}
  \vspace{0.1cm}
  \label{fig:gw3cases}
\end{figure}

The GW energy density of the DWs in the scaling regime would be proportional to the square of DWs surface tension: $\rho^{DW}_{gw}\sim G\mathcal{A}^2\sigma_{\rm dw}^2$~\cite{Hiramatsu:2012sc}, $\sigma_{\rm dw}=(2\sqrt{2}/3)b^{\frac{3}{2}}\lambda_{S}^{-1}v^3$, the area parameter is estimated to be almost $\mathcal{A}\equiv\xi_{\rm dw}=1.09$ in the scaling regime for our simulations. Fig.~\ref{fig:gw3cases}  indicates that the GW spectrum from DWs is found to be almost flat
in intermediate frequencies between the scale corresponding to the horizon size at the decay time of DWs ($k_H$) and that of the DWs width $k_{\rm dw}$, this is because the DWs reach scaling regime and stay in the regime for a long while. 
 The GW energy density from bubble collisions during the first-order PT is expected to $\rho^{PT}_{gw}\sim \rho_c (H_{PT}/\beta)^2(\alpha/(\alpha+1)^2)$ with the $\rho_c$ being the total energy density when the PT occurs~\cite{Huber:2008hg,Konstandin:2017sat}. 
Fig.~\ref{fig:gw3cases} shows that the amplitude of the GW spectra is much higher when the first-order PT occurs, this is mainly because there are more bubble walls in comparison with DWs in the same Hubble volume. Therefore, the contribution to GWs would be dominated by first-order PT rather than the DW 
for the scenarios under study. If the DWs annihilation is driven by a much lower scale of first-order PT, one might comparable contributions of $\rho^{PT}_{gw}\sim\rho^{DW}_{gw}$ which yields $G\mathcal{A}^2\sigma_{dw}^2\sim \left(3H_{PT}^2/(8\pi G)\right) (H_{PT}/\beta)^2\left(\alpha/(\alpha+1)^2\right) $. 
 We numerically confirm that the case with fewer vacuum bubbles can produce a larger amount of GW radiations since it yields a slower PT with larger $H_{PT}/\beta$~\cite{Huber:2008hg,Di:2020kbw,Zhao:2022cnn,Cutting:2018tjt,Cutting:2020nla}, the ratio of the amplitude of the GWs spectra is numerically checked to be: $\Omega_{GW}(N_b=512)/ \Omega_{GW}(N_b=64)\sim ( (H_{PT}/\beta)_{N_b=64}/(H_{PT}/\beta)_{N_b=512})^2 \sim 4$. We therefore can expect that the GWs spectrum generated by long-live DWs decay before BBN driven by the bias term as in literatures~\cite{Hiramatsu:2010yz,Kawasaki:2011vv,Hiramatsu:2013qaa} would be pretty different from that driven by the first-order PT.
Though we cannot obtain the power-law of $k^3$ for the GWs spectra in the IR regions required by the causality due to the limitation of the simulation box, we take $k^3$ for the GWs spectra prediction in the following. The power-law on the right-hand side of the peak frequencies of $k_{\rm H}$ and $k_{\rm b}$ are numerically found to be: $\Omega_{GW}^{DW}\propto k^{-2.14}$ and $\Omega_{GW}(N_b=64,512)\propto k^{-1.60,-1.09}$.

\begin{figure}[!htp]
\includegraphics[width=.45\textwidth]{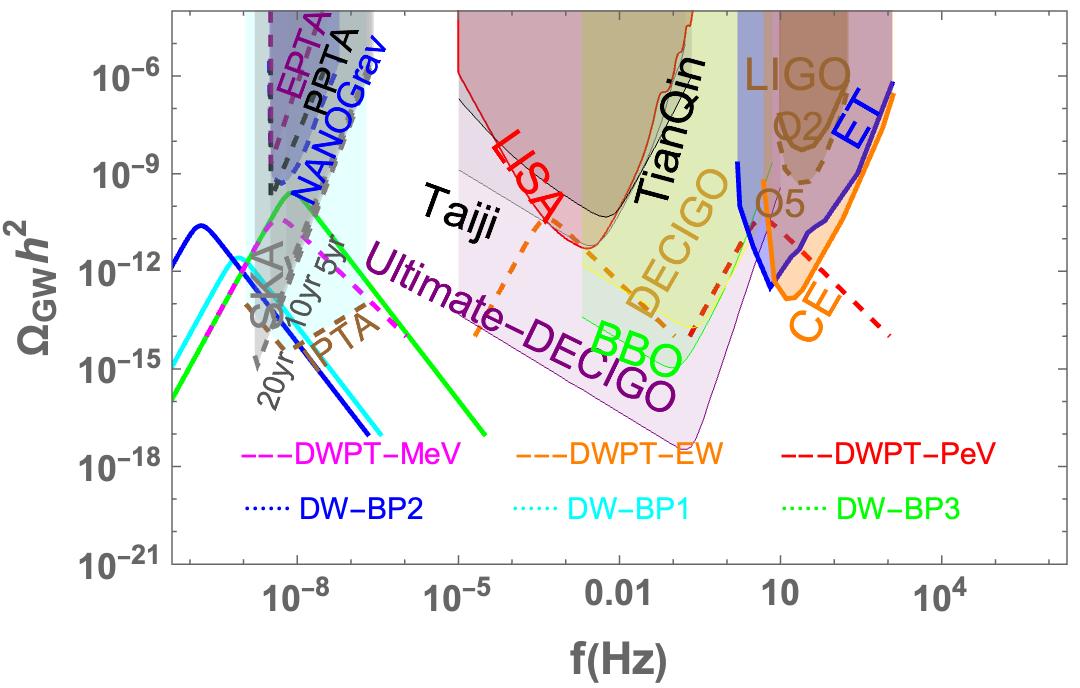} 
  \caption{GWs spectra to be probed at present. The DW-BP1,2,3 corresponds to the discrete symmetry breaking scale $v_s=5,10,100$ TeV and DWs decay temperature $T_{dec}=1,5,50$ MeV.}
  \vspace{0.1cm}
  \label{fig:gws}
\end{figure}

For the GWs generation from the DWs evolution in the scaling regime, we have the peak amplitude being $\Omega^p_{DW}=8\pi \tilde{\epsilon}_{gw}G^2\mathcal{A}^2\sigma^2/(3H^2)$ with the efficiency parameter of GWs production being estimated to be $\tilde{\epsilon}_{gw}=0.49$, see {\it Supplemental material}. Considering the GWs production process ended in the radiation dominated Universe, the present GWs amplitude can be obtained as $\Omega^{p}_{DW}h^2(t_0)=1.67\times 10^{-5}(100/g_\star)^{1/3} \Omega^p_{DW}$ with the relativistic degrees of freedom being $g_\star=106.5$~\cite{Kamionkowski:1993fg}. The present peak frequency determined by the DWs decay before BBN driven by some bias terms can be obtained as~\cite{Saikawa:2017hiv}: $f^p_{DW}(t_0)\approx 1.1\times 10^{-9}  {\rm Hz} (g_\star(T_{dec})/10)^{1/2}(g_{\star,s}(T_{dec})/10)^{-1/3}(T_{dec}/10^{-2} {\rm GeV})$ with $T_{dec}$ denotes the DWs decay temperature.
 For the cases with first-order PT, after considering the scale factor's evolution, the peak amplitude of the present GW spectrum can be estimated as $\Omega^p_{PT}h^2(t_0)=1.67\times 10^{-5}(100/g_\star)^{1/3} \Omega^p_{PT}$ with $\Omega^p_{PT}=(H_{PT}/\beta)^2(\alpha/(\alpha+1)^2)$ at the peak frequency $f^p_{PT}=2.62\times 10^{-3} {\rm mHz} (\beta/H_{PT})(T_{PT}/100 {\rm GeV}) (g_\star/100)^{1/6}$~\cite{Huber:2008hg}.
  For the GWs detection prospects, we only need to keep the GWs spectra around the low-frequency peaks since they have higher magnitudes. In Fig.~\ref{fig:gws}, we present the GWs spectra for DWs and DWs decay by the first-order PT at different scales, with the PeV scale PT can be reached by LIGO~\cite{LIGOScientific:2016aoc,LIGOScientific:2019vic},
Einstein Telescope~\cite{Punturo:2010zz}, and Cosmic Explorer~\cite{Reitze:2019iox}, EW scale PT can be probed by LISA ~\cite{LISA:2017pwj},TianQin~\cite{TianQin:2020hid}, Taiji~\cite{Hu:2017mde,Ruan:2018tsw}, BBO~\cite{Cutler:2005qq}, and DECIGO~\cite{Kudoh:2005as}. MeV-scale PT and DWs at Nano-hertz can be detected by European Pulsar
Timing Array (EPTA~\cite{Desvignes:2016yex}), the Parkes Pulsar Timing Array
(PPTA~\cite{Hobbs:2013aka}), the International Pulsar Timing Array (IPTA
~\cite{Verbiest:2016vem}), the NANOGrav~\cite{NANOGRAV:2018hou}, and SKA~\cite{Janssen:2014dka}.

\noindent{\it \bfseries  Conclusion and discussion.}
The DWs and first-order PTs are generally predicted in particle physics models beyond the Standard Model, and both of them are known to generate SGWB sources. 
In this Letter, we numerically study the phase transition dynamics with both formation of DWs and its decay driven by the following first-order PT. For the first time, our numerical results show that the DWs network can be diluted or annihilated through the interplay between DWs and the vacuum bubbles during the first-order PT. The decay speed of the DWs network is found to be positively correlated with the speed of the first-order PT, i.e., the scaling parameters of the DWs network decrease and drop to zero faster for faster first-order PT. 
We numerically prove that in comparison with the first-order PT, The evolution of the DWs network produces a much lower amount of GWs through the second-order PT is much earlier than the following first-order PT. The SGWB generated by the DWs followed by first-order PT can be probed by future GW detectors, with the MeV-scale scenario has the chance to explain the common red process observed by the NANOGrav 12.5-yr dataset~\cite{Bian:2020urb}.  

In comparison with previous numerical studies of GW spectrum from bubbles collision~\cite{Kosowsky:1992vn,Huber:2008hg,Di:2020kbw,Zhao:2022cnn,Cutting:2018tjt,Cutting:2020nla}, we perform the lattice simulation with more Hubble volume and take into account the scaling parameter and the expansion of the Universe in the radiation dominant Universe. Typically, for the numerical study of GWs from the first-order PT, the hydrodynamics can contribute to the sound wave and are necessary~\cite{Cutting:2019zws,Hindmarsh:2017gnf,Hindmarsh:2013xza,Hindmarsh:2015qta,Konstandin:2017sat,Jinno:2020eqg,Jinno:2022mie}, we left that to future study.


\noindent{\it \bfseries Acknowledgements.}We are grateful to Adrien Florio and Daniel G. Figueroa for helpful discussions on their public code ${\mathcal CosmoLattice}$. The numerical calculations in this study were carried out on the ORISE Supercomputer.
This work is supported in part by the National Key Research and Development Program of China Grants No. 2021YFC2203004, and in part by the National Natural Science Foundation of China under grants Nos. 12075041 and 12147102,  and Chongqing Natural Science Foundation (Grants No.cstc2020jcyj-msxmX0814).

\bibliographystyle{apsrev}

\bibliography{references}

\begin{thebibliography}{96}
\expandafter\ifx\csname natexlab\endcsname\relax\def\natexlab#1{#1}\fi
\expandafter\ifx\csname bibnamefont\endcsname\relax
  \def\bibnamefont#1{#1}\fi
\expandafter\ifx\csname bibfnamefont\endcsname\relax
  \def\bibfnamefont#1{#1}\fi
\expandafter\ifx\csname citenamefont\endcsname\relax
  \def\citenamefont#1{#1}\fi
\expandafter\ifx\csname url\endcsname\relax
  \def\url#1{\texttt{#1}}\fi
\expandafter\ifx\csname urlprefix\endcsname\relax\def\urlprefix{URL }\fi
\providecommand{\bibinfo}[2]{#2}
\providecommand{\eprint}[2][]{\url{#2}}

\bibitem[{\citenamefont{Kibble}(1976)}]{kibble1976topology}
\bibinfo{author}{\bibfnamefont{T.~W.~B.} \bibnamefont{Kibble}},
  \bibinfo{journal}{J. Phys. A} \textbf{\bibinfo{volume}{9}},
  \bibinfo{pages}{1387} (\bibinfo{year}{1976}).

\bibitem[{\citenamefont{Zurek}(1985)}]{zurek1985cosmological}
\bibinfo{author}{\bibfnamefont{W.~H.} \bibnamefont{Zurek}},
  \bibinfo{journal}{Nature} \textbf{\bibinfo{volume}{317}},
  \bibinfo{pages}{505} (\bibinfo{year}{1985}).

\bibitem[{\citenamefont{Zeldovich et~al.}(1974)\citenamefont{Zeldovich,
  Kobzarev, and Okun}}]{Zeldovich:1974uw}
\bibinfo{author}{\bibfnamefont{Y.~B.} \bibnamefont{Zeldovich}},
  \bibinfo{author}{\bibfnamefont{I.~Y.} \bibnamefont{Kobzarev}},
  \bibnamefont{and} \bibinfo{author}{\bibfnamefont{L.~B.} \bibnamefont{Okun}},
  \bibinfo{journal}{Zh. Eksp. Teor. Fiz.} \textbf{\bibinfo{volume}{67}},
  \bibinfo{pages}{3} (\bibinfo{year}{1974}).

\bibitem[{\citenamefont{Sikivie}(1982{\natexlab{a}})}]{Sikivie:1982qv}
\bibinfo{author}{\bibfnamefont{P.}~\bibnamefont{Sikivie}},
  \bibinfo{journal}{Phys. Rev. Lett.} \textbf{\bibinfo{volume}{48}},
  \bibinfo{pages}{1156} (\bibinfo{year}{1982}{\natexlab{a}}).

\bibitem[{\citenamefont{Vilenkin and Everett}(1982)}]{Vilenkin:1982ks}
\bibinfo{author}{\bibfnamefont{A.}~\bibnamefont{Vilenkin}} \bibnamefont{and}
  \bibinfo{author}{\bibfnamefont{A.~E.} \bibnamefont{Everett}},
  \bibinfo{journal}{Phys. Rev. Lett.} \textbf{\bibinfo{volume}{48}},
  \bibinfo{pages}{1867} (\bibinfo{year}{1982}).

\bibitem[{\citenamefont{Hiramatsu et~al.}(2011)\citenamefont{Hiramatsu,
  Kawasaki, and Saikawa}}]{Hiramatsu:2010yn}
\bibinfo{author}{\bibfnamefont{T.}~\bibnamefont{Hiramatsu}},
  \bibinfo{author}{\bibfnamefont{M.}~\bibnamefont{Kawasaki}}, \bibnamefont{and}
  \bibinfo{author}{\bibfnamefont{K.}~\bibnamefont{Saikawa}},
  \bibinfo{journal}{JCAP} \textbf{\bibinfo{volume}{08}}, \bibinfo{pages}{030}
  (\bibinfo{year}{2011}), \eprint{1012.4558}.

\bibitem[{\citenamefont{Hiramatsu et~al.}(2013)\citenamefont{Hiramatsu,
  Kawasaki, Saikawa, and Sekiguchi}}]{Hiramatsu:2012sc}
\bibinfo{author}{\bibfnamefont{T.}~\bibnamefont{Hiramatsu}},
  \bibinfo{author}{\bibfnamefont{M.}~\bibnamefont{Kawasaki}},
  \bibinfo{author}{\bibfnamefont{K.}~\bibnamefont{Saikawa}}, \bibnamefont{and}
  \bibinfo{author}{\bibfnamefont{T.}~\bibnamefont{Sekiguchi}},
  \bibinfo{journal}{JCAP} \textbf{\bibinfo{volume}{01}}, \bibinfo{pages}{001}
  (\bibinfo{year}{2013}), \eprint{1207.3166}.

\bibitem[{\citenamefont{Kawasaki et~al.}(2015)\citenamefont{Kawasaki, Saikawa,
  and Sekiguchi}}]{Kawasaki:2014sqa}
\bibinfo{author}{\bibfnamefont{M.}~\bibnamefont{Kawasaki}},
  \bibinfo{author}{\bibfnamefont{K.}~\bibnamefont{Saikawa}}, \bibnamefont{and}
  \bibinfo{author}{\bibfnamefont{T.}~\bibnamefont{Sekiguchi}},
  \bibinfo{journal}{Phys. Rev. D} \textbf{\bibinfo{volume}{91}},
  \bibinfo{pages}{065014} (\bibinfo{year}{2015}), \eprint{1412.0789}.

\bibitem[{\citenamefont{Abel et~al.}(1995)\citenamefont{Abel, Sarkar, and
  White}}]{Abel:1995wk}
\bibinfo{author}{\bibfnamefont{S.~A.} \bibnamefont{Abel}},
  \bibinfo{author}{\bibfnamefont{S.}~\bibnamefont{Sarkar}}, \bibnamefont{and}
  \bibinfo{author}{\bibfnamefont{P.~L.} \bibnamefont{White}},
  \bibinfo{journal}{Nucl. Phys. B} \textbf{\bibinfo{volume}{454}},
  \bibinfo{pages}{663} (\bibinfo{year}{1995}), \eprint{hep-ph/9506359}.

\bibitem[{\citenamefont{Takahashi et~al.}(2008)\citenamefont{Takahashi,
  Yanagida, and Yonekura}}]{Takahashi:2008mu}
\bibinfo{author}{\bibfnamefont{F.}~\bibnamefont{Takahashi}},
  \bibinfo{author}{\bibfnamefont{T.~T.} \bibnamefont{Yanagida}},
  \bibnamefont{and} \bibinfo{author}{\bibfnamefont{K.}~\bibnamefont{Yonekura}},
  \bibinfo{journal}{Phys. Lett. B} \textbf{\bibinfo{volume}{664}},
  \bibinfo{pages}{194} (\bibinfo{year}{2008}), \eprint{0802.4335}.

\bibitem[{\citenamefont{Dine et~al.}(2010)\citenamefont{Dine, Takahashi, and
  Yanagida}}]{Dine:2010eb}
\bibinfo{author}{\bibfnamefont{M.}~\bibnamefont{Dine}},
  \bibinfo{author}{\bibfnamefont{F.}~\bibnamefont{Takahashi}},
  \bibnamefont{and} \bibinfo{author}{\bibfnamefont{T.~T.}
  \bibnamefont{Yanagida}}, \bibinfo{journal}{JHEP}
  \textbf{\bibinfo{volume}{07}}, \bibinfo{pages}{003} (\bibinfo{year}{2010}),
  \eprint{1005.3613}.

\bibitem[{\citenamefont{Hamaguchi et~al.}(2012)\citenamefont{Hamaguchi,
  Nakayama, and Yokozaki}}]{Hamaguchi:2011nm}
\bibinfo{author}{\bibfnamefont{K.}~\bibnamefont{Hamaguchi}},
  \bibinfo{author}{\bibfnamefont{K.}~\bibnamefont{Nakayama}}, \bibnamefont{and}
  \bibinfo{author}{\bibfnamefont{N.}~\bibnamefont{Yokozaki}},
  \bibinfo{journal}{Phys. Lett. B} \textbf{\bibinfo{volume}{708}},
  \bibinfo{pages}{100} (\bibinfo{year}{2012}), \eprint{1107.4760}.

\bibitem[{\citenamefont{Kadota et~al.}(2015)\citenamefont{Kadota, Kawasaki, and
  Saikawa}}]{Kadota:2015dza}
\bibinfo{author}{\bibfnamefont{K.}~\bibnamefont{Kadota}},
  \bibinfo{author}{\bibfnamefont{M.}~\bibnamefont{Kawasaki}}, \bibnamefont{and}
  \bibinfo{author}{\bibfnamefont{K.}~\bibnamefont{Saikawa}},
  \bibinfo{journal}{JCAP} \textbf{\bibinfo{volume}{10}}, \bibinfo{pages}{041}
  (\bibinfo{year}{2015}), \eprint{1503.06998}.

\bibitem[{\citenamefont{Moroi and Nakayama}(2011)}]{Moroi:2011be}
\bibinfo{author}{\bibfnamefont{T.}~\bibnamefont{Moroi}} \bibnamefont{and}
  \bibinfo{author}{\bibfnamefont{K.}~\bibnamefont{Nakayama}},
  \bibinfo{journal}{Phys. Lett. B} \textbf{\bibinfo{volume}{703}},
  \bibinfo{pages}{160} (\bibinfo{year}{2011}), \eprint{1105.6216}.

\bibitem[{\citenamefont{Coulson et~al.}(1996)\citenamefont{Coulson, Lalak, and
  Ovrut}}]{Coulson:1995nv}
\bibinfo{author}{\bibfnamefont{D.}~\bibnamefont{Coulson}},
  \bibinfo{author}{\bibfnamefont{Z.}~\bibnamefont{Lalak}}, \bibnamefont{and}
  \bibinfo{author}{\bibfnamefont{B.~A.} \bibnamefont{Ovrut}},
  \bibinfo{journal}{Phys. Rev. D} \textbf{\bibinfo{volume}{53}},
  \bibinfo{pages}{4237} (\bibinfo{year}{1996}).

\bibitem[{\citenamefont{Gelmini et~al.}(1989)\citenamefont{Gelmini, Gleiser,
  and Kolb}}]{Gelmini:1988sf}
\bibinfo{author}{\bibfnamefont{G.~B.} \bibnamefont{Gelmini}},
  \bibinfo{author}{\bibfnamefont{M.}~\bibnamefont{Gleiser}}, \bibnamefont{and}
  \bibinfo{author}{\bibfnamefont{E.~W.} \bibnamefont{Kolb}},
  \bibinfo{journal}{Phys. Rev. D} \textbf{\bibinfo{volume}{39}},
  \bibinfo{pages}{1558} (\bibinfo{year}{1989}).

\bibitem[{\citenamefont{Vilenkin}(1981)}]{Vilenkin:1981zs}
\bibinfo{author}{\bibfnamefont{A.}~\bibnamefont{Vilenkin}},
  \bibinfo{journal}{Phys. Rev. D} \textbf{\bibinfo{volume}{23}},
  \bibinfo{pages}{852} (\bibinfo{year}{1981}).

\bibitem[{\citenamefont{Larsson et~al.}(1997)\citenamefont{Larsson, Sarkar, and
  White}}]{Larsson:1996sp}
\bibinfo{author}{\bibfnamefont{S.~E.} \bibnamefont{Larsson}},
  \bibinfo{author}{\bibfnamefont{S.}~\bibnamefont{Sarkar}}, \bibnamefont{and}
  \bibinfo{author}{\bibfnamefont{P.~L.} \bibnamefont{White}},
  \bibinfo{journal}{Phys. Rev. D} \textbf{\bibinfo{volume}{55}},
  \bibinfo{pages}{5129} (\bibinfo{year}{1997}), \eprint{hep-ph/9608319}.

\bibitem[{\citenamefont{Sikivie}(1982{\natexlab{b}})}]{sikivie1982axions}
\bibinfo{author}{\bibfnamefont{P.}~\bibnamefont{Sikivie}},
  \bibinfo{journal}{Phys. Rev. Lett.} \textbf{\bibinfo{volume}{48}},
  \bibinfo{pages}{1156} (\bibinfo{year}{1982}{\natexlab{b}}).

\bibitem[{\citenamefont{Gleiser and Roberts}(1998)}]{Gleiser:1998na}
\bibinfo{author}{\bibfnamefont{M.}~\bibnamefont{Gleiser}} \bibnamefont{and}
  \bibinfo{author}{\bibfnamefont{R.}~\bibnamefont{Roberts}},
  \bibinfo{journal}{Phys. Rev. Lett.} \textbf{\bibinfo{volume}{81}},
  \bibinfo{pages}{5497} (\bibinfo{year}{1998}), \eprint{astro-ph/9807260}.

\bibitem[{\citenamefont{Hiramatsu et~al.}(2010)\citenamefont{Hiramatsu,
  Kawasaki, and Saikawa}}]{Hiramatsu:2010yz}
\bibinfo{author}{\bibfnamefont{T.}~\bibnamefont{Hiramatsu}},
  \bibinfo{author}{\bibfnamefont{M.}~\bibnamefont{Kawasaki}}, \bibnamefont{and}
  \bibinfo{author}{\bibfnamefont{K.}~\bibnamefont{Saikawa}},
  \bibinfo{journal}{JCAP} \textbf{\bibinfo{volume}{05}}, \bibinfo{pages}{032}
  (\bibinfo{year}{2010}), \eprint{1002.1555}.

\bibitem[{\citenamefont{Kawasaki and Saikawa}(2011)}]{Kawasaki:2011vv}
\bibinfo{author}{\bibfnamefont{M.}~\bibnamefont{Kawasaki}} \bibnamefont{and}
  \bibinfo{author}{\bibfnamefont{K.}~\bibnamefont{Saikawa}},
  \bibinfo{journal}{JCAP} \textbf{\bibinfo{volume}{09}}, \bibinfo{pages}{008}
  (\bibinfo{year}{2011}), \eprint{1102.5628}.

\bibitem[{\citenamefont{Hiramatsu et~al.}(2014)\citenamefont{Hiramatsu,
  Kawasaki, and Saikawa}}]{Hiramatsu:2013qaa}
\bibinfo{author}{\bibfnamefont{T.}~\bibnamefont{Hiramatsu}},
  \bibinfo{author}{\bibfnamefont{M.}~\bibnamefont{Kawasaki}}, \bibnamefont{and}
  \bibinfo{author}{\bibfnamefont{K.}~\bibnamefont{Saikawa}},
  \bibinfo{journal}{JCAP} \textbf{\bibinfo{volume}{02}}, \bibinfo{pages}{031}
  (\bibinfo{year}{2014}), \eprint{1309.5001}.

\bibitem[{\citenamefont{Bian et~al.}(2022)\citenamefont{Bian, Ge, Li, Shu, and
  Zong}}]{Bian:2022qbh}
\bibinfo{author}{\bibfnamefont{L.}~\bibnamefont{Bian}},
  \bibinfo{author}{\bibfnamefont{S.}~\bibnamefont{Ge}},
  \bibinfo{author}{\bibfnamefont{C.}~\bibnamefont{Li}},
  \bibinfo{author}{\bibfnamefont{J.}~\bibnamefont{Shu}}, \bibnamefont{and}
  \bibinfo{author}{\bibfnamefont{J.}~\bibnamefont{Zong}}
  (\bibinfo{year}{2022}), \eprint{2212.07871}.

\bibitem[{\citenamefont{Ferreira et~al.}(2023)\citenamefont{Ferreira, Notari,
  Pujolas, and Rompineve}}]{Ferreira:2022zzo}
\bibinfo{author}{\bibfnamefont{R.~Z.} \bibnamefont{Ferreira}},
  \bibinfo{author}{\bibfnamefont{A.}~\bibnamefont{Notari}},
  \bibinfo{author}{\bibfnamefont{O.}~\bibnamefont{Pujolas}}, \bibnamefont{and}
  \bibinfo{author}{\bibfnamefont{F.}~\bibnamefont{Rompineve}},
  \bibinfo{journal}{JCAP} \textbf{\bibinfo{volume}{02}}, \bibinfo{pages}{001}
  (\bibinfo{year}{2023}), \eprint{2204.04228}.

\bibitem[{\citenamefont{Caprini et~al.}(2016)}]{Caprini:2015zlo}
\bibinfo{author}{\bibfnamefont{C.}~\bibnamefont{Caprini}} \bibnamefont{et~al.},
  \bibinfo{journal}{JCAP} \textbf{\bibinfo{volume}{04}}, \bibinfo{pages}{001}
  (\bibinfo{year}{2016}), \eprint{1512.06239}.

\bibitem[{\citenamefont{Bian et~al.}(2021{\natexlab{a}})}]{Bian:2021ini}
\bibinfo{author}{\bibfnamefont{L.}~\bibnamefont{Bian}} \bibnamefont{et~al.},
  \bibinfo{journal}{Sci. China Phys. Mech. Astron.}
  \textbf{\bibinfo{volume}{64}}, \bibinfo{pages}{120401}
  (\bibinfo{year}{2021}{\natexlab{a}}), \eprint{2106.10235}.

\bibitem[{\citenamefont{Caprini et~al.}(2020)}]{Caprini:2019egz}
\bibinfo{author}{\bibfnamefont{C.}~\bibnamefont{Caprini}} \bibnamefont{et~al.},
  \bibinfo{journal}{JCAP} \textbf{\bibinfo{volume}{03}}, \bibinfo{pages}{024}
  (\bibinfo{year}{2020}), \eprint{1910.13125}.

\bibitem[{\citenamefont{Cai et~al.}(2017)\citenamefont{Cai, Cao, Guo, Wang, and
  Yang}}]{Cai:2017cbj}
\bibinfo{author}{\bibfnamefont{R.-G.} \bibnamefont{Cai}},
  \bibinfo{author}{\bibfnamefont{Z.}~\bibnamefont{Cao}},
  \bibinfo{author}{\bibfnamefont{Z.-K.} \bibnamefont{Guo}},
  \bibinfo{author}{\bibfnamefont{S.-J.} \bibnamefont{Wang}}, \bibnamefont{and}
  \bibinfo{author}{\bibfnamefont{T.}~\bibnamefont{Yang}},
  \bibinfo{journal}{Natl. Sci. Rev.} \textbf{\bibinfo{volume}{4}},
  \bibinfo{pages}{687} (\bibinfo{year}{2017}), \eprint{1703.00187}.

\bibitem[{\citenamefont{Caldwell et~al.}(2022)}]{Caldwell:2022qsj}
\bibinfo{author}{\bibfnamefont{R.}~\bibnamefont{Caldwell}}
  \bibnamefont{et~al.}, \bibinfo{journal}{Gen. Rel. Grav.}
  \textbf{\bibinfo{volume}{54}}, \bibinfo{pages}{156} (\bibinfo{year}{2022}),
  \eprint{2203.07972}.

\bibitem[{\citenamefont{Bian et~al.}(2021{\natexlab{b}})\citenamefont{Bian,
  Cai, Liu, Yang, and Zhou}}]{Bian:2020urb}
\bibinfo{author}{\bibfnamefont{L.}~\bibnamefont{Bian}},
  \bibinfo{author}{\bibfnamefont{R.-G.} \bibnamefont{Cai}},
  \bibinfo{author}{\bibfnamefont{J.}~\bibnamefont{Liu}},
  \bibinfo{author}{\bibfnamefont{X.-Y.} \bibnamefont{Yang}}, \bibnamefont{and}
  \bibinfo{author}{\bibfnamefont{R.}~\bibnamefont{Zhou}},
  \bibinfo{journal}{Phys. Rev. D} \textbf{\bibinfo{volume}{103}},
  \bibinfo{pages}{L081301} (\bibinfo{year}{2021}{\natexlab{b}}),
  \eprint{2009.13893}.

\bibitem[{\citenamefont{Xue et~al.}(2021)}]{Xue:2021gyq}
\bibinfo{author}{\bibfnamefont{X.}~\bibnamefont{Xue}} \bibnamefont{et~al.},
  \bibinfo{journal}{Phys. Rev. Lett.} \textbf{\bibinfo{volume}{127}},
  \bibinfo{pages}{251303} (\bibinfo{year}{2021}), \eprint{2110.03096}.

\bibitem[{\citenamefont{Arzoumanian et~al.}(2021)}]{NANOGrav:2021flc}
\bibinfo{author}{\bibfnamefont{Z.}~\bibnamefont{Arzoumanian}}
  \bibnamefont{et~al.} (\bibinfo{collaboration}{NANOGrav}),
  \bibinfo{journal}{Phys. Rev. Lett.} \textbf{\bibinfo{volume}{127}},
  \bibinfo{pages}{251302} (\bibinfo{year}{2021}), \eprint{2104.13930}.

\bibitem[{\citenamefont{Romero et~al.}(2021)\citenamefont{Romero, Martinovic,
  Callister, Guo, Mart\'\i{}nez, Sakellariadou, Yang, and
  Zhao}}]{Romero:2021kby}
\bibinfo{author}{\bibfnamefont{A.}~\bibnamefont{Romero}},
  \bibinfo{author}{\bibfnamefont{K.}~\bibnamefont{Martinovic}},
  \bibinfo{author}{\bibfnamefont{T.~A.} \bibnamefont{Callister}},
  \bibinfo{author}{\bibfnamefont{H.-K.} \bibnamefont{Guo}},
  \bibinfo{author}{\bibfnamefont{M.}~\bibnamefont{Mart\'\i{}nez}},
  \bibinfo{author}{\bibfnamefont{M.}~\bibnamefont{Sakellariadou}},
  \bibinfo{author}{\bibfnamefont{F.-W.} \bibnamefont{Yang}}, \bibnamefont{and}
  \bibinfo{author}{\bibfnamefont{Y.}~\bibnamefont{Zhao}},
  \bibinfo{journal}{Phys. Rev. Lett.} \textbf{\bibinfo{volume}{126}},
  \bibinfo{pages}{151301} (\bibinfo{year}{2021}), \eprint{2102.01714}.

\bibitem[{\citenamefont{McDonald}(1994)}]{McDonald:1993ex}
\bibinfo{author}{\bibfnamefont{J.}~\bibnamefont{McDonald}},
  \bibinfo{journal}{Phys. Rev. D} \textbf{\bibinfo{volume}{50}},
  \bibinfo{pages}{3637} (\bibinfo{year}{1994}), \eprint{hep-ph/0702143}.

\bibitem[{\citenamefont{Burgess et~al.}(2001)\citenamefont{Burgess, Pospelov,
  and ter Veldhuis}}]{Burgess:2000yq}
\bibinfo{author}{\bibfnamefont{C.~P.} \bibnamefont{Burgess}},
  \bibinfo{author}{\bibfnamefont{M.}~\bibnamefont{Pospelov}}, \bibnamefont{and}
  \bibinfo{author}{\bibfnamefont{T.}~\bibnamefont{ter Veldhuis}},
  \bibinfo{journal}{Nucl. Phys. B} \textbf{\bibinfo{volume}{619}},
  \bibinfo{pages}{709} (\bibinfo{year}{2001}), \eprint{hep-ph/0011335}.

\bibitem[{\citenamefont{Espinosa and Quiros}(2007)}]{Espinosa:2007qk}
\bibinfo{author}{\bibfnamefont{J.~R.} \bibnamefont{Espinosa}} \bibnamefont{and}
  \bibinfo{author}{\bibfnamefont{M.}~\bibnamefont{Quiros}},
  \bibinfo{journal}{Phys. Rev. D} \textbf{\bibinfo{volume}{76}},
  \bibinfo{pages}{076004} (\bibinfo{year}{2007}), \eprint{hep-ph/0701145}.

\bibitem[{\citenamefont{Profumo et~al.}(2007)\citenamefont{Profumo,
  Ramsey-Musolf, and Shaughnessy}}]{Profumo:2007wc}
\bibinfo{author}{\bibfnamefont{S.}~\bibnamefont{Profumo}},
  \bibinfo{author}{\bibfnamefont{M.~J.} \bibnamefont{Ramsey-Musolf}},
  \bibnamefont{and}
  \bibinfo{author}{\bibfnamefont{G.}~\bibnamefont{Shaughnessy}},
  \bibinfo{journal}{JHEP} \textbf{\bibinfo{volume}{08}}, \bibinfo{pages}{010}
  (\bibinfo{year}{2007}), \eprint{0705.2425}.

\bibitem[{\citenamefont{Barger et~al.}(2008)\citenamefont{Barger, Langacker,
  McCaskey, Ramsey-Musolf, and Shaughnessy}}]{Barger:2007im}
\bibinfo{author}{\bibfnamefont{V.}~\bibnamefont{Barger}},
  \bibinfo{author}{\bibfnamefont{P.}~\bibnamefont{Langacker}},
  \bibinfo{author}{\bibfnamefont{M.}~\bibnamefont{McCaskey}},
  \bibinfo{author}{\bibfnamefont{M.~J.} \bibnamefont{Ramsey-Musolf}},
  \bibnamefont{and}
  \bibinfo{author}{\bibfnamefont{G.}~\bibnamefont{Shaughnessy}},
  \bibinfo{journal}{Phys. Rev. D} \textbf{\bibinfo{volume}{77}},
  \bibinfo{pages}{035005} (\bibinfo{year}{2008}), \eprint{0706.4311}.

\bibitem[{\citenamefont{Espinosa et~al.}(2008)\citenamefont{Espinosa,
  Konstandin, No, and Quiros}}]{Espinosa:2008kw}
\bibinfo{author}{\bibfnamefont{J.~R.} \bibnamefont{Espinosa}},
  \bibinfo{author}{\bibfnamefont{T.}~\bibnamefont{Konstandin}},
  \bibinfo{author}{\bibfnamefont{J.~M.} \bibnamefont{No}}, \bibnamefont{and}
  \bibinfo{author}{\bibfnamefont{M.}~\bibnamefont{Quiros}},
  \bibinfo{journal}{Phys. Rev. D} \textbf{\bibinfo{volume}{78}},
  \bibinfo{pages}{123528} (\bibinfo{year}{2008}), \eprint{0809.3215}.

\bibitem[{\citenamefont{Espinosa
  et~al.}(2012{\natexlab{a}})\citenamefont{Espinosa, Konstandin, and
  Riva}}]{Espinosa:2011ax}
\bibinfo{author}{\bibfnamefont{J.~R.} \bibnamefont{Espinosa}},
  \bibinfo{author}{\bibfnamefont{T.}~\bibnamefont{Konstandin}},
  \bibnamefont{and} \bibinfo{author}{\bibfnamefont{F.}~\bibnamefont{Riva}},
  \bibinfo{journal}{Nucl. Phys. B} \textbf{\bibinfo{volume}{854}},
  \bibinfo{pages}{592} (\bibinfo{year}{2012}{\natexlab{a}}),
  \eprint{1107.5441}.

\bibitem[{\citenamefont{Espinosa
  et~al.}(2012{\natexlab{b}})\citenamefont{Espinosa, Gripaios, Konstandin, and
  Riva}}]{Espinosa:2011eu}
\bibinfo{author}{\bibfnamefont{J.~R.} \bibnamefont{Espinosa}},
  \bibinfo{author}{\bibfnamefont{B.}~\bibnamefont{Gripaios}},
  \bibinfo{author}{\bibfnamefont{T.}~\bibnamefont{Konstandin}},
  \bibnamefont{and} \bibinfo{author}{\bibfnamefont{F.}~\bibnamefont{Riva}},
  \bibinfo{journal}{JCAP} \textbf{\bibinfo{volume}{01}}, \bibinfo{pages}{012}
  (\bibinfo{year}{2012}{\natexlab{b}}), \eprint{1110.2876}.

\bibitem[{\citenamefont{Cline and Kainulainen}(2013)}]{Cline:2012hg}
\bibinfo{author}{\bibfnamefont{J.~M.} \bibnamefont{Cline}} \bibnamefont{and}
  \bibinfo{author}{\bibfnamefont{K.}~\bibnamefont{Kainulainen}},
  \bibinfo{journal}{JCAP} \textbf{\bibinfo{volume}{01}}, \bibinfo{pages}{012}
  (\bibinfo{year}{2013}), \eprint{1210.4196}.

\bibitem[{\citenamefont{Profumo et~al.}(2015)\citenamefont{Profumo,
  Ramsey-Musolf, Wainwright, and Winslow}}]{Profumo:2014opa}
\bibinfo{author}{\bibfnamefont{S.}~\bibnamefont{Profumo}},
  \bibinfo{author}{\bibfnamefont{M.~J.} \bibnamefont{Ramsey-Musolf}},
  \bibinfo{author}{\bibfnamefont{C.~L.} \bibnamefont{Wainwright}},
  \bibnamefont{and} \bibinfo{author}{\bibfnamefont{P.}~\bibnamefont{Winslow}},
  \bibinfo{journal}{Phys. Rev. D} \textbf{\bibinfo{volume}{91}},
  \bibinfo{pages}{035018} (\bibinfo{year}{2015}), \eprint{1407.5342}.

\bibitem[{\citenamefont{Feng et~al.}(2015)\citenamefont{Feng, Profumo, and
  Ubaldi}}]{Feng:2014vea}
\bibinfo{author}{\bibfnamefont{L.}~\bibnamefont{Feng}},
  \bibinfo{author}{\bibfnamefont{S.}~\bibnamefont{Profumo}}, \bibnamefont{and}
  \bibinfo{author}{\bibfnamefont{L.}~\bibnamefont{Ubaldi}},
  \bibinfo{journal}{JHEP} \textbf{\bibinfo{volume}{03}}, \bibinfo{pages}{045}
  (\bibinfo{year}{2015}), \eprint{1412.1105}.

\bibitem[{\citenamefont{Curtin et~al.}(2014)\citenamefont{Curtin, Meade, and
  Yu}}]{Curtin:2014jma}
\bibinfo{author}{\bibfnamefont{D.}~\bibnamefont{Curtin}},
  \bibinfo{author}{\bibfnamefont{P.}~\bibnamefont{Meade}}, \bibnamefont{and}
  \bibinfo{author}{\bibfnamefont{C.-T.} \bibnamefont{Yu}},
  \bibinfo{journal}{JHEP} \textbf{\bibinfo{volume}{11}}, \bibinfo{pages}{127}
  (\bibinfo{year}{2014}), \eprint{1409.0005}.

\bibitem[{\citenamefont{Craig et~al.}(2016)\citenamefont{Craig, Lou,
  McCullough, and Thalapillil}}]{Craig:2014lda}
\bibinfo{author}{\bibfnamefont{N.}~\bibnamefont{Craig}},
  \bibinfo{author}{\bibfnamefont{H.~K.} \bibnamefont{Lou}},
  \bibinfo{author}{\bibfnamefont{M.}~\bibnamefont{McCullough}},
  \bibnamefont{and}
  \bibinfo{author}{\bibfnamefont{A.}~\bibnamefont{Thalapillil}},
  \bibinfo{journal}{JHEP} \textbf{\bibinfo{volume}{02}}, \bibinfo{pages}{127}
  (\bibinfo{year}{2016}), \eprint{1412.0258}.

\bibitem[{\citenamefont{Huang et~al.}(2016)\citenamefont{Huang, Long, and
  Wang}}]{Huang:2016cjm}
\bibinfo{author}{\bibfnamefont{P.}~\bibnamefont{Huang}},
  \bibinfo{author}{\bibfnamefont{A.~J.} \bibnamefont{Long}}, \bibnamefont{and}
  \bibinfo{author}{\bibfnamefont{L.-T.} \bibnamefont{Wang}},
  \bibinfo{journal}{Phys. Rev. D} \textbf{\bibinfo{volume}{94}},
  \bibinfo{pages}{075008} (\bibinfo{year}{2016}), \eprint{1608.06619}.

\bibitem[{\citenamefont{Vaskonen}(2017)}]{Vaskonen:2016yiu}
\bibinfo{author}{\bibfnamefont{V.}~\bibnamefont{Vaskonen}},
  \bibinfo{journal}{Phys. Rev. D} \textbf{\bibinfo{volume}{95}},
  \bibinfo{pages}{123515} (\bibinfo{year}{2017}), \eprint{1611.02073}.

\bibitem[{\citenamefont{Curtin et~al.}(2018)\citenamefont{Curtin, Meade, and
  Ramani}}]{Curtin:2016urg}
\bibinfo{author}{\bibfnamefont{D.}~\bibnamefont{Curtin}},
  \bibinfo{author}{\bibfnamefont{P.}~\bibnamefont{Meade}}, \bibnamefont{and}
  \bibinfo{author}{\bibfnamefont{H.}~\bibnamefont{Ramani}},
  \bibinfo{journal}{Eur. Phys. J. C} \textbf{\bibinfo{volume}{78}},
  \bibinfo{pages}{787} (\bibinfo{year}{2018}), \eprint{1612.00466}.

\bibitem[{\citenamefont{Kurup and Perelstein}(2017)}]{Kurup:2017dzf}
\bibinfo{author}{\bibfnamefont{G.}~\bibnamefont{Kurup}} \bibnamefont{and}
  \bibinfo{author}{\bibfnamefont{M.}~\bibnamefont{Perelstein}},
  \bibinfo{journal}{Phys. Rev. D} \textbf{\bibinfo{volume}{96}},
  \bibinfo{pages}{015036} (\bibinfo{year}{2017}), \eprint{1704.03381}.

\bibitem[{\citenamefont{Buttazzo et~al.}(2018)\citenamefont{Buttazzo, Redigolo,
  Sala, and Tesi}}]{Buttazzo:2018qqp}
\bibinfo{author}{\bibfnamefont{D.}~\bibnamefont{Buttazzo}},
  \bibinfo{author}{\bibfnamefont{D.}~\bibnamefont{Redigolo}},
  \bibinfo{author}{\bibfnamefont{F.}~\bibnamefont{Sala}}, \bibnamefont{and}
  \bibinfo{author}{\bibfnamefont{A.}~\bibnamefont{Tesi}},
  \bibinfo{journal}{JHEP} \textbf{\bibinfo{volume}{11}}, \bibinfo{pages}{144}
  (\bibinfo{year}{2018}), \eprint{1807.04743}.

\bibitem[{\citenamefont{Alanne et~al.}(2020)\citenamefont{Alanne, Hugle,
  Platscher, and Schmitz}}]{Alanne:2019bsm}
\bibinfo{author}{\bibfnamefont{T.}~\bibnamefont{Alanne}},
  \bibinfo{author}{\bibfnamefont{T.}~\bibnamefont{Hugle}},
  \bibinfo{author}{\bibfnamefont{M.}~\bibnamefont{Platscher}},
  \bibnamefont{and} \bibinfo{author}{\bibfnamefont{K.}~\bibnamefont{Schmitz}},
  \bibinfo{journal}{JHEP} \textbf{\bibinfo{volume}{03}}, \bibinfo{pages}{004}
  (\bibinfo{year}{2020}), \eprint{1909.11356}.

\bibitem[{\citenamefont{Costantini et~al.}(2020)\citenamefont{Costantini,
  De~Lillo, Maltoni, Mantani, Mattelaer, Ruiz, and Zhao}}]{Costantini:2020stv}
\bibinfo{author}{\bibfnamefont{A.}~\bibnamefont{Costantini}},
  \bibinfo{author}{\bibfnamefont{F.}~\bibnamefont{De~Lillo}},
  \bibinfo{author}{\bibfnamefont{F.}~\bibnamefont{Maltoni}},
  \bibinfo{author}{\bibfnamefont{L.}~\bibnamefont{Mantani}},
  \bibinfo{author}{\bibfnamefont{O.}~\bibnamefont{Mattelaer}},
  \bibinfo{author}{\bibfnamefont{R.}~\bibnamefont{Ruiz}}, \bibnamefont{and}
  \bibinfo{author}{\bibfnamefont{X.}~\bibnamefont{Zhao}},
  \bibinfo{journal}{JHEP} \textbf{\bibinfo{volume}{09}}, \bibinfo{pages}{080}
  (\bibinfo{year}{2020}), \eprint{2005.10289}.

\bibitem[{\citenamefont{Al~Ali et~al.}(2022)}]{AlAli:2021let}
\bibinfo{author}{\bibfnamefont{H.}~\bibnamefont{Al~Ali}} \bibnamefont{et~al.},
  \bibinfo{journal}{Rept. Prog. Phys.} \textbf{\bibinfo{volume}{85}},
  \bibinfo{pages}{084201} (\bibinfo{year}{2022}), \eprint{2103.14043}.

\bibitem[{\citenamefont{McDonald}(1995)}]{McDonald:1995hp}
\bibinfo{author}{\bibfnamefont{J.}~\bibnamefont{McDonald}},
  \bibinfo{journal}{Phys. Lett. B} \textbf{\bibinfo{volume}{357}},
  \bibinfo{pages}{19} (\bibinfo{year}{1995}).

\bibitem[{\citenamefont{Blasi and Mariotti}(2022)}]{Blasi:2022woz}
\bibinfo{author}{\bibfnamefont{S.}~\bibnamefont{Blasi}} \bibnamefont{and}
  \bibinfo{author}{\bibfnamefont{A.}~\bibnamefont{Mariotti}},
  \bibinfo{journal}{Phys. Rev. Lett.} \textbf{\bibinfo{volume}{129}},
  \bibinfo{pages}{261303} (\bibinfo{year}{2022}), \eprint{2203.16450}.

\bibitem[{\citenamefont{Blasi et~al.}(2023)\citenamefont{Blasi, Jinno,
  Konstandin, Rubira, and Stomberg}}]{Blasi:2023rqi}
\bibinfo{author}{\bibfnamefont{S.}~\bibnamefont{Blasi}},
  \bibinfo{author}{\bibfnamefont{R.}~\bibnamefont{Jinno}},
  \bibinfo{author}{\bibfnamefont{T.}~\bibnamefont{Konstandin}},
  \bibinfo{author}{\bibfnamefont{H.}~\bibnamefont{Rubira}}, \bibnamefont{and}
  \bibinfo{author}{\bibfnamefont{I.}~\bibnamefont{Stomberg}}
  (\bibinfo{year}{2023}), \eprint{2302.06952}.

\bibitem[{\citenamefont{Bian and Tang}(2018)}]{Bian:2018mkl}
\bibinfo{author}{\bibfnamefont{L.}~\bibnamefont{Bian}} \bibnamefont{and}
  \bibinfo{author}{\bibfnamefont{Y.-L.} \bibnamefont{Tang}},
  \bibinfo{journal}{JHEP} \textbf{\bibinfo{volume}{12}}, \bibinfo{pages}{006}
  (\bibinfo{year}{2018}), \eprint{1810.03172}.

\bibitem[{\citenamefont{Figueroa et~al.}(2023)\citenamefont{Figueroa, Florio,
  Torrenti, and Valkenburg}}]{Figueroa:2021yhd}
\bibinfo{author}{\bibfnamefont{D.~G.} \bibnamefont{Figueroa}},
  \bibinfo{author}{\bibfnamefont{A.}~\bibnamefont{Florio}},
  \bibinfo{author}{\bibfnamefont{F.}~\bibnamefont{Torrenti}}, \bibnamefont{and}
  \bibinfo{author}{\bibfnamefont{W.}~\bibnamefont{Valkenburg}},
  \bibinfo{journal}{Comput. Phys. Commun.} \textbf{\bibinfo{volume}{283}},
  \bibinfo{pages}{108586} (\bibinfo{year}{2023}), \eprint{2102.01031}.

\bibitem[{\citenamefont{Figueroa et~al.}(2021)\citenamefont{Figueroa, Florio,
  Torrenti, and Valkenburg}}]{Figueroa:2020rrl}
\bibinfo{author}{\bibfnamefont{D.~G.} \bibnamefont{Figueroa}},
  \bibinfo{author}{\bibfnamefont{A.}~\bibnamefont{Florio}},
  \bibinfo{author}{\bibfnamefont{F.}~\bibnamefont{Torrenti}}, \bibnamefont{and}
  \bibinfo{author}{\bibfnamefont{W.}~\bibnamefont{Valkenburg}},
  \bibinfo{journal}{JCAP} \textbf{\bibinfo{volume}{04}}, \bibinfo{pages}{035}
  (\bibinfo{year}{2021}), \eprint{2006.15122}.

\bibitem[{\citenamefont{Hindmarsh et~al.}(2021)\citenamefont{Hindmarsh,
  L\"uben, Lumma, and Pauly}}]{Hindmarsh:2020hop}
\bibinfo{author}{\bibfnamefont{M.~B.} \bibnamefont{Hindmarsh}},
  \bibinfo{author}{\bibfnamefont{M.}~\bibnamefont{L\"uben}},
  \bibinfo{author}{\bibfnamefont{J.}~\bibnamefont{Lumma}}, \bibnamefont{and}
  \bibinfo{author}{\bibfnamefont{M.}~\bibnamefont{Pauly}},
  \bibinfo{journal}{SciPost Phys. Lect. Notes} \textbf{\bibinfo{volume}{24}},
  \bibinfo{pages}{1} (\bibinfo{year}{2021}), \eprint{2008.09136}.

\bibitem[{\citenamefont{Huber and Konstandin}(2008)}]{Huber:2008hg}
\bibinfo{author}{\bibfnamefont{S.~J.} \bibnamefont{Huber}} \bibnamefont{and}
  \bibinfo{author}{\bibfnamefont{T.}~\bibnamefont{Konstandin}},
  \bibinfo{journal}{JCAP} \textbf{\bibinfo{volume}{09}}, \bibinfo{pages}{022}
  (\bibinfo{year}{2008}), \eprint{0806.1828}.

\bibitem[{\citenamefont{Konstandin}(2018)}]{Konstandin:2017sat}
\bibinfo{author}{\bibfnamefont{T.}~\bibnamefont{Konstandin}},
  \bibinfo{journal}{JCAP} \textbf{\bibinfo{volume}{03}}, \bibinfo{pages}{047}
  (\bibinfo{year}{2018}), \eprint{1712.06869}.

\bibitem[{\citenamefont{Di et~al.}(2021)\citenamefont{Di, Wang, Zhou, Bian,
  Cai, and Liu}}]{Di:2020kbw}
\bibinfo{author}{\bibfnamefont{Y.}~\bibnamefont{Di}},
  \bibinfo{author}{\bibfnamefont{J.}~\bibnamefont{Wang}},
  \bibinfo{author}{\bibfnamefont{R.}~\bibnamefont{Zhou}},
  \bibinfo{author}{\bibfnamefont{L.}~\bibnamefont{Bian}},
  \bibinfo{author}{\bibfnamefont{R.-G.} \bibnamefont{Cai}}, \bibnamefont{and}
  \bibinfo{author}{\bibfnamefont{J.}~\bibnamefont{Liu}},
  \bibinfo{journal}{Phys. Rev. Lett.} \textbf{\bibinfo{volume}{126}},
  \bibinfo{pages}{251102} (\bibinfo{year}{2021}), \eprint{2012.15625}.

\bibitem[{\citenamefont{Zhao et~al.}(2022)\citenamefont{Zhao, Di, Bian, and
  Cai}}]{Zhao:2022cnn}
\bibinfo{author}{\bibfnamefont{Z.}~\bibnamefont{Zhao}},
  \bibinfo{author}{\bibfnamefont{Y.}~\bibnamefont{Di}},
  \bibinfo{author}{\bibfnamefont{L.}~\bibnamefont{Bian}}, \bibnamefont{and}
  \bibinfo{author}{\bibfnamefont{R.-G.} \bibnamefont{Cai}}
  (\bibinfo{year}{2022}), \eprint{2204.04427}.

\bibitem[{\citenamefont{Cutting et~al.}(2018)\citenamefont{Cutting, Hindmarsh,
  and Weir}}]{Cutting:2018tjt}
\bibinfo{author}{\bibfnamefont{D.}~\bibnamefont{Cutting}},
  \bibinfo{author}{\bibfnamefont{M.}~\bibnamefont{Hindmarsh}},
  \bibnamefont{and} \bibinfo{author}{\bibfnamefont{D.~J.} \bibnamefont{Weir}},
  \bibinfo{journal}{Phys. Rev. D} \textbf{\bibinfo{volume}{97}},
  \bibinfo{pages}{123513} (\bibinfo{year}{2018}), \eprint{1802.05712}.

\bibitem[{\citenamefont{Cutting et~al.}(2021)\citenamefont{Cutting, Escartin,
  Hindmarsh, and Weir}}]{Cutting:2020nla}
\bibinfo{author}{\bibfnamefont{D.}~\bibnamefont{Cutting}},
  \bibinfo{author}{\bibfnamefont{E.~G.} \bibnamefont{Escartin}},
  \bibinfo{author}{\bibfnamefont{M.}~\bibnamefont{Hindmarsh}},
  \bibnamefont{and} \bibinfo{author}{\bibfnamefont{D.~J.} \bibnamefont{Weir}},
  \bibinfo{journal}{Phys. Rev. D} \textbf{\bibinfo{volume}{103}},
  \bibinfo{pages}{023531} (\bibinfo{year}{2021}), \eprint{2005.13537}.

\bibitem[{\citenamefont{Kamionkowski et~al.}(1994)\citenamefont{Kamionkowski,
  Kosowsky, and Turner}}]{Kamionkowski:1993fg}
\bibinfo{author}{\bibfnamefont{M.}~\bibnamefont{Kamionkowski}},
  \bibinfo{author}{\bibfnamefont{A.}~\bibnamefont{Kosowsky}}, \bibnamefont{and}
  \bibinfo{author}{\bibfnamefont{M.~S.} \bibnamefont{Turner}},
  \bibinfo{journal}{Phys. Rev. D} \textbf{\bibinfo{volume}{49}},
  \bibinfo{pages}{2837} (\bibinfo{year}{1994}), \eprint{astro-ph/9310044}.

\bibitem[{\citenamefont{Saikawa}(2017)}]{Saikawa:2017hiv}
\bibinfo{author}{\bibfnamefont{K.}~\bibnamefont{Saikawa}},
  \bibinfo{journal}{Universe} \textbf{\bibinfo{volume}{3}}, \bibinfo{pages}{40}
  (\bibinfo{year}{2017}), \eprint{1703.02576}.

\bibitem[{\citenamefont{Abbott et~al.}(2016)}]{LIGOScientific:2016aoc}
\bibinfo{author}{\bibfnamefont{B.~P.} \bibnamefont{Abbott}}
  \bibnamefont{et~al.} (\bibinfo{collaboration}{LIGO Scientific, Virgo}),
  \bibinfo{journal}{Phys. Rev. Lett.} \textbf{\bibinfo{volume}{116}},
  \bibinfo{pages}{061102} (\bibinfo{year}{2016}), \eprint{1602.03837}.

\bibitem[{\citenamefont{Abbott et~al.}(2019)}]{LIGOScientific:2019vic}
\bibinfo{author}{\bibfnamefont{B.~P.} \bibnamefont{Abbott}}
  \bibnamefont{et~al.} (\bibinfo{collaboration}{LIGO Scientific, Virgo}),
  \bibinfo{journal}{Phys. Rev. D} \textbf{\bibinfo{volume}{100}},
  \bibinfo{pages}{061101} (\bibinfo{year}{2019}), \eprint{1903.02886}.

\bibitem[{\citenamefont{Punturo et~al.}(2010)}]{Punturo:2010zz}
\bibinfo{author}{\bibfnamefont{M.}~\bibnamefont{Punturo}} \bibnamefont{et~al.},
  \bibinfo{journal}{Class. Quant. Grav.} \textbf{\bibinfo{volume}{27}},
  \bibinfo{pages}{194002} (\bibinfo{year}{2010}).

\bibitem[{\citenamefont{Reitze et~al.}(2019)}]{Reitze:2019iox}
\bibinfo{author}{\bibfnamefont{D.}~\bibnamefont{Reitze}} \bibnamefont{et~al.},
  \bibinfo{journal}{Bull. Am. Astron. Soc.} \textbf{\bibinfo{volume}{51}},
  \bibinfo{pages}{035} (\bibinfo{year}{2019}), \eprint{1907.04833}.

\bibitem[{\citenamefont{Amaro-Seoane et~al.}(2017)}]{LISA:2017pwj}
\bibinfo{author}{\bibfnamefont{P.}~\bibnamefont{Amaro-Seoane}}
  \bibnamefont{et~al.} (\bibinfo{collaboration}{LISA}) (\bibinfo{year}{2017}),
  \eprint{1702.00786}.

\bibitem[{\citenamefont{Mei et~al.}(2021)}]{TianQin:2020hid}
\bibinfo{author}{\bibfnamefont{J.}~\bibnamefont{Mei}} \bibnamefont{et~al.}
  (\bibinfo{collaboration}{TianQin}), \bibinfo{journal}{PTEP}
  \textbf{\bibinfo{volume}{2021}}, \bibinfo{pages}{05A107}
  (\bibinfo{year}{2021}), \eprint{2008.10332}.

\bibitem[{\citenamefont{Hu and Wu}(2017)}]{Hu:2017mde}
\bibinfo{author}{\bibfnamefont{W.-R.} \bibnamefont{Hu}} \bibnamefont{and}
  \bibinfo{author}{\bibfnamefont{Y.-L.} \bibnamefont{Wu}},
  \bibinfo{journal}{Natl. Sci. Rev.} \textbf{\bibinfo{volume}{4}},
  \bibinfo{pages}{685} (\bibinfo{year}{2017}).

\bibitem[{\citenamefont{Ruan et~al.}(2020)\citenamefont{Ruan, Guo, Cai, and
  Zhang}}]{Ruan:2018tsw}
\bibinfo{author}{\bibfnamefont{W.-H.} \bibnamefont{Ruan}},
  \bibinfo{author}{\bibfnamefont{Z.-K.} \bibnamefont{Guo}},
  \bibinfo{author}{\bibfnamefont{R.-G.} \bibnamefont{Cai}}, \bibnamefont{and}
  \bibinfo{author}{\bibfnamefont{Y.-Z.} \bibnamefont{Zhang}},
  \bibinfo{journal}{Int. J. Mod. Phys. A} \textbf{\bibinfo{volume}{35}},
  \bibinfo{pages}{2050075} (\bibinfo{year}{2020}), \eprint{1807.09495}.

\bibitem[{\citenamefont{Cutler and Harms}(2006)}]{Cutler:2005qq}
\bibinfo{author}{\bibfnamefont{C.}~\bibnamefont{Cutler}} \bibnamefont{and}
  \bibinfo{author}{\bibfnamefont{J.}~\bibnamefont{Harms}},
  \bibinfo{journal}{Phys. Rev. D} \textbf{\bibinfo{volume}{73}},
  \bibinfo{pages}{042001} (\bibinfo{year}{2006}), \eprint{gr-qc/0511092}.

\bibitem[{\citenamefont{Kudoh et~al.}(2006)\citenamefont{Kudoh, Taruya,
  Hiramatsu, and Himemoto}}]{Kudoh:2005as}
\bibinfo{author}{\bibfnamefont{H.}~\bibnamefont{Kudoh}},
  \bibinfo{author}{\bibfnamefont{A.}~\bibnamefont{Taruya}},
  \bibinfo{author}{\bibfnamefont{T.}~\bibnamefont{Hiramatsu}},
  \bibnamefont{and} \bibinfo{author}{\bibfnamefont{Y.}~\bibnamefont{Himemoto}},
  \bibinfo{journal}{Phys. Rev. D} \textbf{\bibinfo{volume}{73}},
  \bibinfo{pages}{064006} (\bibinfo{year}{2006}), \eprint{gr-qc/0511145}.

\bibitem[{\citenamefont{Desvignes et~al.}(2016)}]{Desvignes:2016yex}
\bibinfo{author}{\bibfnamefont{G.}~\bibnamefont{Desvignes}}
  \bibnamefont{et~al.}, \bibinfo{journal}{Mon. Not. Roy. Astron. Soc.}
  \textbf{\bibinfo{volume}{458}}, \bibinfo{pages}{3341} (\bibinfo{year}{2016}),
  \eprint{1602.08511}.

\bibitem[{\citenamefont{Hobbs}(2013)}]{Hobbs:2013aka}
\bibinfo{author}{\bibfnamefont{G.}~\bibnamefont{Hobbs}},
  \bibinfo{journal}{Class. Quant. Grav.} \textbf{\bibinfo{volume}{30}},
  \bibinfo{pages}{224007} (\bibinfo{year}{2013}), \eprint{1307.2629}.

\bibitem[{\citenamefont{Verbiest et~al.}(2016)}]{Verbiest:2016vem}
\bibinfo{author}{\bibfnamefont{J.~P.~W.} \bibnamefont{Verbiest}}
  \bibnamefont{et~al.}, \bibinfo{journal}{Mon. Not. Roy. Astron. Soc.}
  \textbf{\bibinfo{volume}{458}}, \bibinfo{pages}{1267} (\bibinfo{year}{2016}),
  \eprint{1602.03640}.

\bibitem[{\citenamefont{Arzoumanian et~al.}(2018)}]{NANOGRAV:2018hou}
\bibinfo{author}{\bibfnamefont{Z.}~\bibnamefont{Arzoumanian}}
  \bibnamefont{et~al.} (\bibinfo{collaboration}{NANOGRAV}),
  \bibinfo{journal}{Astrophys. J.} \textbf{\bibinfo{volume}{859}},
  \bibinfo{pages}{47} (\bibinfo{year}{2018}), \eprint{1801.02617}.

\bibitem[{\citenamefont{Janssen et~al.}(2015)}]{Janssen:2014dka}
\bibinfo{author}{\bibfnamefont{G.}~\bibnamefont{Janssen}} \bibnamefont{et~al.},
  \bibinfo{journal}{PoS} \textbf{\bibinfo{volume}{AASKA14}},
  \bibinfo{pages}{037} (\bibinfo{year}{2015}), \eprint{1501.00127}.

\bibitem[{\citenamefont{Kosowsky and Turner}(1993)}]{Kosowsky:1992vn}
\bibinfo{author}{\bibfnamefont{A.}~\bibnamefont{Kosowsky}} \bibnamefont{and}
  \bibinfo{author}{\bibfnamefont{M.~S.} \bibnamefont{Turner}},
  \bibinfo{journal}{Phys. Rev. D} \textbf{\bibinfo{volume}{47}},
  \bibinfo{pages}{4372} (\bibinfo{year}{1993}), \eprint{astro-ph/9211004}.

\bibitem[{\citenamefont{Cutting et~al.}(2020)\citenamefont{Cutting, Hindmarsh,
  and Weir}}]{Cutting:2019zws}
\bibinfo{author}{\bibfnamefont{D.}~\bibnamefont{Cutting}},
  \bibinfo{author}{\bibfnamefont{M.}~\bibnamefont{Hindmarsh}},
  \bibnamefont{and} \bibinfo{author}{\bibfnamefont{D.~J.} \bibnamefont{Weir}},
  \bibinfo{journal}{Phys. Rev. Lett.} \textbf{\bibinfo{volume}{125}},
  \bibinfo{pages}{021302} (\bibinfo{year}{2020}), \eprint{1906.00480}.

\bibitem[{\citenamefont{Hindmarsh et~al.}(2017)\citenamefont{Hindmarsh, Huber,
  Rummukainen, and Weir}}]{Hindmarsh:2017gnf}
\bibinfo{author}{\bibfnamefont{M.}~\bibnamefont{Hindmarsh}},
  \bibinfo{author}{\bibfnamefont{S.~J.} \bibnamefont{Huber}},
  \bibinfo{author}{\bibfnamefont{K.}~\bibnamefont{Rummukainen}},
  \bibnamefont{and} \bibinfo{author}{\bibfnamefont{D.~J.} \bibnamefont{Weir}},
  \bibinfo{journal}{Phys. Rev. D} \textbf{\bibinfo{volume}{96}},
  \bibinfo{pages}{103520} (\bibinfo{year}{2017}), \bibinfo{note}{[Erratum:
  Phys.Rev.D 101, 089902 (2020)]}, \eprint{1704.05871}.

\bibitem[{\citenamefont{Hindmarsh et~al.}(2014)\citenamefont{Hindmarsh, Huber,
  Rummukainen, and Weir}}]{Hindmarsh:2013xza}
\bibinfo{author}{\bibfnamefont{M.}~\bibnamefont{Hindmarsh}},
  \bibinfo{author}{\bibfnamefont{S.~J.} \bibnamefont{Huber}},
  \bibinfo{author}{\bibfnamefont{K.}~\bibnamefont{Rummukainen}},
  \bibnamefont{and} \bibinfo{author}{\bibfnamefont{D.~J.} \bibnamefont{Weir}},
  \bibinfo{journal}{Phys. Rev. Lett.} \textbf{\bibinfo{volume}{112}},
  \bibinfo{pages}{041301} (\bibinfo{year}{2014}), \eprint{1304.2433}.

\bibitem[{\citenamefont{Hindmarsh et~al.}(2015)\citenamefont{Hindmarsh, Huber,
  Rummukainen, and Weir}}]{Hindmarsh:2015qta}
\bibinfo{author}{\bibfnamefont{M.}~\bibnamefont{Hindmarsh}},
  \bibinfo{author}{\bibfnamefont{S.~J.} \bibnamefont{Huber}},
  \bibinfo{author}{\bibfnamefont{K.}~\bibnamefont{Rummukainen}},
  \bibnamefont{and} \bibinfo{author}{\bibfnamefont{D.~J.} \bibnamefont{Weir}},
  \bibinfo{journal}{Phys. Rev. D} \textbf{\bibinfo{volume}{92}},
  \bibinfo{pages}{123009} (\bibinfo{year}{2015}), \eprint{1504.03291}.

\bibitem[{\citenamefont{Jinno et~al.}(2021)\citenamefont{Jinno, Konstandin, and
  Rubira}}]{Jinno:2020eqg}
\bibinfo{author}{\bibfnamefont{R.}~\bibnamefont{Jinno}},
  \bibinfo{author}{\bibfnamefont{T.}~\bibnamefont{Konstandin}},
  \bibnamefont{and} \bibinfo{author}{\bibfnamefont{H.}~\bibnamefont{Rubira}},
  \bibinfo{journal}{JCAP} \textbf{\bibinfo{volume}{04}}, \bibinfo{pages}{014}
  (\bibinfo{year}{2021}), \eprint{2010.00971}.

\bibitem[{\citenamefont{Jinno et~al.}(2023)\citenamefont{Jinno, Konstandin,
  Rubira, and Stomberg}}]{Jinno:2022mie}
\bibinfo{author}{\bibfnamefont{R.}~\bibnamefont{Jinno}},
  \bibinfo{author}{\bibfnamefont{T.}~\bibnamefont{Konstandin}},
  \bibinfo{author}{\bibfnamefont{H.}~\bibnamefont{Rubira}}, \bibnamefont{and}
  \bibinfo{author}{\bibfnamefont{I.}~\bibnamefont{Stomberg}},
  \bibinfo{journal}{JCAP} \textbf{\bibinfo{volume}{02}}, \bibinfo{pages}{011}
  (\bibinfo{year}{2023}), \eprint{2209.04369}.

\bibitem[{\citenamefont{Press et~al.}(1989)\citenamefont{Press, Ryden, and
  Spergel}}]{Press:1989yh}
\bibinfo{author}{\bibfnamefont{W.~H.} \bibnamefont{Press}},
  \bibinfo{author}{\bibfnamefont{B.~S.} \bibnamefont{Ryden}}, \bibnamefont{and}
  \bibinfo{author}{\bibfnamefont{D.~N.} \bibnamefont{Spergel}},
  \bibinfo{journal}{Astrophys. J.} \textbf{\bibinfo{volume}{347}},
  \bibinfo{pages}{590} (\bibinfo{year}{1989}).

\bibitem[{\citenamefont{Dufaux et~al.}(2007)\citenamefont{Dufaux, Bergman,
  Felder, Kofman, and Uzan}}]{Dufaux:2007pt}
\bibinfo{author}{\bibfnamefont{J.~F.} \bibnamefont{Dufaux}},
  \bibinfo{author}{\bibfnamefont{A.}~\bibnamefont{Bergman}},
  \bibinfo{author}{\bibfnamefont{G.~N.} \bibnamefont{Felder}},
  \bibinfo{author}{\bibfnamefont{L.}~\bibnamefont{Kofman}}, \bibnamefont{and}
  \bibinfo{author}{\bibfnamefont{J.-P.} \bibnamefont{Uzan}},
  \bibinfo{journal}{Phys. Rev. D} \textbf{\bibinfo{volume}{76}},
  \bibinfo{pages}{123517} (\bibinfo{year}{2007}), \eprint{0707.0875}.

\bibitem[{\citenamefont{Easther et~al.}(2008)\citenamefont{Easther, Giblin, and
  Lim}}]{Easther:2007vj}
\bibinfo{author}{\bibfnamefont{R.}~\bibnamefont{Easther}},
  \bibinfo{author}{\bibfnamefont{J.~T.} \bibnamefont{Giblin}},
  \bibnamefont{and} \bibinfo{author}{\bibfnamefont{E.~A.} \bibnamefont{Lim}},
  \bibinfo{journal}{Phys. Rev. D} \textbf{\bibinfo{volume}{77}},
  \bibinfo{pages}{103519} (\bibinfo{year}{2008}), \eprint{0712.2991}.

\bibitem[{\citenamefont{Price and Siemens}(2008)}]{Price:2008hq}
\bibinfo{author}{\bibfnamefont{L.~R.} \bibnamefont{Price}} \bibnamefont{and}
  \bibinfo{author}{\bibfnamefont{X.}~\bibnamefont{Siemens}},
  \bibinfo{journal}{Phys. Rev. D} \textbf{\bibinfo{volume}{78}},
  \bibinfo{pages}{063541} (\bibinfo{year}{2008}), \eprint{0805.3570}.

\end{thebibliography}

\clearpage

\onecolumngrid
\begin{center}
  \textbf{\large Supplementary Material}\\[.2cm]
\end{center}

\onecolumngrid
\setcounter{equation}{0}
\setcounter{figure}{0}
\setcounter{table}{0}
\setcounter{section}{0}
\setcounter{page}{1}
\makeatletter
\renewcommand{\theequation}{S\arabic{equation}}
\renewcommand{\thefigure}{S\arabic{figure}}

This supplementary material contains the details of our simulation and interpretation of results presented in the main text, as well as some additional results. We start with a detailed explanation of our numerical scheme, including equations of motion and initial conditions. Next, we introduce the methods of identifying the domain wall on the lattice, together with the calculation of the scaling parameter. Then, we explain the detailed procedures for calculating the energy density power spectrum of gravitational waves. At last, we presented the results of our numerical simulations.

\section{Equations of motion}\label{sec:SupEOM}
\begin{figure*}[htp]
\includegraphics[width=.33\textwidth]{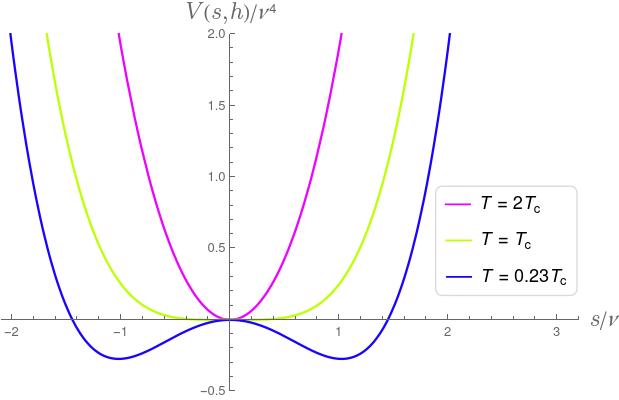} 
\hspace{3mm}
\includegraphics[width=.32\textwidth]{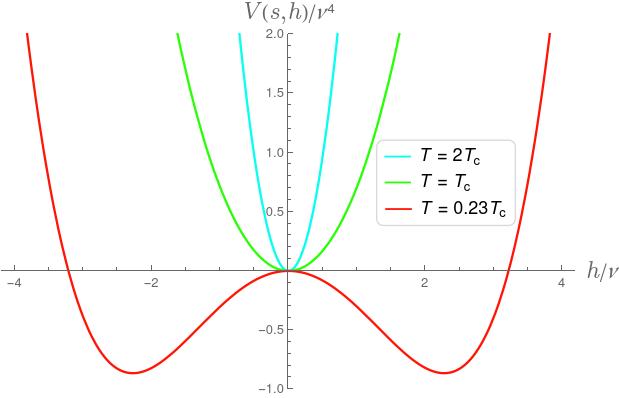}
\hspace{3mm}
\includegraphics[width=.25\textwidth]{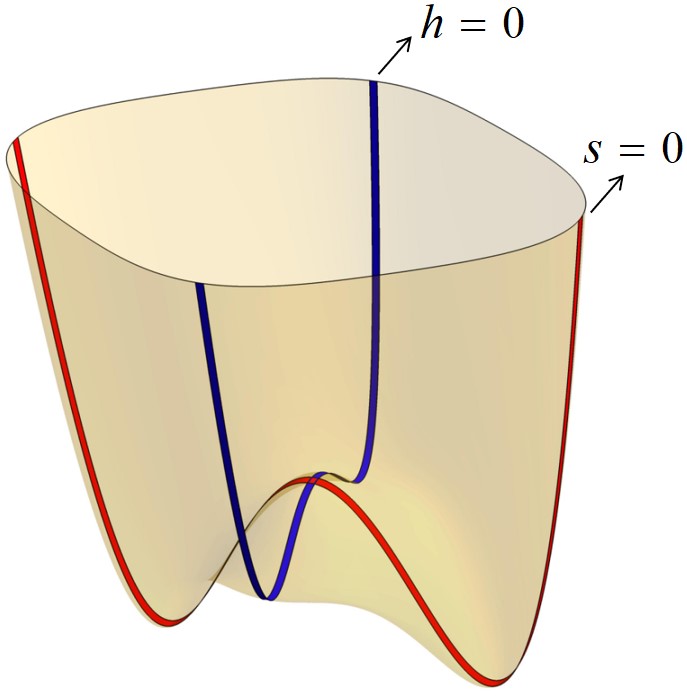}
  \caption{The shape of potential at different temperatures, where $T_{\rm c}=2.17v$ is the critical temperature of second-order phase transition, $T=0.23T_{\rm c}$ is the temperature at which bubbles are generated in the simulation box. The left panel shows the shape of potential in the $s$ direction ($h=0$), the middle panel shows the shape of potential in the $h$ direction ($s=0$), and the right panel shows the three-dimensional distribution of potential near the origin. }
  \vspace{0.1cm}
  \label{fig:shape of the potential}
\end{figure*}

We perform lattice simulations in a radiation-dominated universe with spatially flat FLRW metric
\begin{equation}
{\rm d}s^2=g_{\mu\nu}{\rm d}x^{\mu}{\rm d}x^{\nu}=a^2(\eta)(-{\rm d}\eta^2+{\rm d}\boldsymbol{x}^2),
\end{equation}
where $\eta=\int{\frac{{\rm d}t}{a(t)}}$ is the conformal time.
The field equations of motion can be obtained by varying this action
\begin{equation}
S=-\int{d^4x\sqrt{-g}\left\{ \frac{1}{2}\partial_\mu s \partial^\mu s + \frac{1}{2}\partial_\mu h \partial^\mu h + V(s,h)) \right\}},
\end{equation}
where $s$ and $h$ are two singlet scalar. 
The shape of the potential energy as a function of temperature is shown in Fig.~\ref{fig:shape of the potential}. 

For the convenience of numerical simulation, we introduce two parameters $f_*$ and $w_*$ to do the rescale for physical variables
\begin{equation}
\tilde{s}=s/f_*, \ \tilde{h}=h/f_*, \ {\rm d}\tilde{\eta}={\rm d}\eta w_*=\frac{1}{a}w_*{\rm d}t, \ {\rm d}\tilde{x}^i=w_*{\rm d}x^i, \ \tilde{V}(\tilde{s},\tilde{h})=\frac{V(f_*\tilde{s},f_*\tilde{h})}{f_*^2 w_*^2}, \ \tilde{S}=(\frac{w_*}{f_*})^2S.  
\end{equation}
In all of our simulations, we set $f_*=v$ and $w_*=a_iH_i$, where $a_i$ and $H_i$ are the initial scale factor (we set $a_i=0.5$) and Hubble parameter, respectively. By varying the dimensionless action $\tilde{S}$, we can obtain the equation of motion expressed by dimensionless fields and space-time variables  
\begin{eqnarray}
&&\tilde{s}''+2\frac{a'}{a}\tilde{s}'-\tilde{\nabla}^2 \tilde{s} =-a^2\frac{\partial \tilde{V}(\tilde{s},\tilde{h})}{\partial \tilde{s}}, \\
&&\tilde{h}''+2\frac{a'}{a}\tilde{h}'-\tilde{\nabla}^2 \tilde{h} =-a^2\frac{\partial \tilde{V}(\tilde{s},\tilde{h})}{\partial \tilde{h}},
\end{eqnarray}
with $'={\rm d}/{\rm d}\tilde{\eta}$ and $\tilde{\nabla}_i={\rm d}/{\rm d}\tilde{x}^i$.

To reduce the number of iterations required to solve equations of motion, we use the rescaled conformal time $\hat{\eta}$ in our simulations
\begin{equation}
\hat{\eta}=\frac{\eta}{\eta_i}=\frac{a(\eta)}{a_i}=\frac{T_i}{T(\eta)}=\left(\frac{t}{t_i}\right)^{1/2},
\end{equation}
where the physical quantities with "$i$" in the subscript are defined at the initial time. The choice of the initial conformal time ($\eta_{i}$) is not arbitrary, since it is related to the evolution of the scale factor. The evolution of the scale factor in the radiation-dominated universe is determined by the solution of Friedmann equations
\begin{equation}
a(\eta)=a_i\left(1+ \frac{1}{p}a_iH_i\times(\eta-\eta_i) \right)^p, \quad \text{with} \ p \equiv \frac{2}{3(1+w)-2} \ \text{and} \ w=\frac{1}{3}.
\end{equation}
To make the scale factor meet the conditions in Eq.(S6) at the same time, the initial conformal time must be chosen as $\eta_i=1/(a_iH_i)$. Note that $\eta_i=1/w_*$, it is easy to see that the rescaled conformal time is actually the dimensionless conformal time (see Eq.(S3)), $\hat{\eta}=\tilde{\eta}$, so we will directly use $\tilde{\eta}$ instead of $\hat{\eta}$ in the following parts.

\section{Initial Conditions} 
\label{sec:initial conditions}
The physical scenario we investigated experienced a second-order phase transition in the early stage of simulation. Before the second-order phase transition, the temperature is much higher than the critical temperature, and the scalar field $s$ is in thermal equilibrium. Therefore, we can use the thermal spectrum to describe the amplitude and momentum distribution of the scalar field $s$ in momentum space
\begin{equation}
\mathcal{P}_{s}(k)=\frac{n_k}{w_k}=\frac{1}{w_k}\frac{1}{e^{w_k/T}-1}, \quad \mathcal{P}_{\dot{s}}(k)=n_k w_k=\frac{w_k}{e^{w_k/T}-1},
\end{equation}
where $n_k=1/(e^{w_k/T}-1)$ is the occupation number of the Bose-Einstein distribution, $w_k=\sqrt{k^2/a^2+m_{\rm eff}^2}$ and $k$ are physical frequency and comoving momenta, respectively, $m_{\rm eff}$ is the initial effective mass of $s$ and overdots denote differentiation with respect to cosmic time.

In the continuum, correlation functions of the scalar field $s$ in momentum space can be written as
\begin{align}
\langle s (\boldsymbol{{\rm k}}) s(\boldsymbol{{\rm k}}') \rangle &= (2\pi)^3\mathcal{P}_{s}(k)\delta(\boldsymbol{{\rm k}}-\boldsymbol{{\rm k}}'), \\
\langle \dot{s} (\boldsymbol{{\rm k}}) \dot{s}(\boldsymbol{{\rm k}}') \rangle &= (2\pi)^3\mathcal{P}_{\dot{s}}(k)\delta(\boldsymbol{{\rm k}}-\boldsymbol{{\rm k}}'), \\
\langle s(\boldsymbol{{\rm k}}) \dot{s}(\boldsymbol{{\rm k}}') \rangle &= 0,  \ 
\end{align}
where $\boldsymbol{{\rm k}}$ denotes three dimensional momentum in Fourier space and $\langle \cdots \rangle$ represents here an ensemble average.
With appropriate rescaling~\cite{Figueroa:2020rrl}, we can reproduce the correlation functions equivalent to that in the continuum on a discrete lattice, which does not depend explicitly on the volume 
\begin{align}
\qquad \langle | s(\boldsymbol{{\rm k}}) |^2 \rangle &= (\frac{N}{\delta x_{\rm phy}})^3\mathcal{P}_{s}(k), \qquad \langle s(\boldsymbol{{\rm k}})  \rangle = 0, \\
\qquad \langle | \dot s(\boldsymbol{{\rm k}}) |^2 \rangle &= (\frac{N}{\delta x_{\rm phy}})^3\mathcal{P}_{\dot{s}}(k),\qquad \langle \dot s(\boldsymbol{{\rm k}}) \rangle = 0,
\end{align}
with $N$ denote the number of points each side and $\delta x_{\rm phy}$ is the physical lattice spacing. We generate $s(\boldsymbol{{\rm k}})$ and $\dot s(\boldsymbol{{\rm k}})$ following Gaussian random distribution in momentum space, 
and include all modes from infrared truncation ($k_{\rm IR}$) to ultraviolet truncation ($k_{\rm UV}$) of the simulation box. Finally, we can obtain the field in three-dimensional coordinate space by applying discrete Fourier transform (DFT) to the field in momentum space, and the correlation functions in coordinate space are given by 
\begin{align}
\langle s (\boldsymbol{{\rm x}}) s(\boldsymbol{{\rm x}}') \rangle &= \frac{1}{(2\pi)^3}\int {\rm d}^3k\mathcal{P}_{s}(k)e^{-i\boldsymbol{{\rm k}}\cdot (\boldsymbol{{\rm x}}-\boldsymbol{{\rm x}}') }, \\
\langle \dot{s} (\boldsymbol{{\rm x}}) \dot{s}(\boldsymbol{{\rm x}}') \rangle &= \frac{1}{(2\pi)^3}\int {\rm d}^3k\mathcal{P}_{\dot{s}}(k)e^{-i\boldsymbol{{\rm k}}\cdot (\boldsymbol{{\rm x}}-\boldsymbol{{\rm x}}') }, \\
\langle s(\boldsymbol{{\rm x}}) \dot{s}(\boldsymbol{{\rm x}}') \rangle &= 0.  \ 
\end{align}

\section{Identification and scaling parameter of domain wall}

In principle, we can identify where the domain wall exists according to the value of the potential energy since the domain wall exists near the false vacuum with higher potential energy after spontaneous symmetry breaking. But in fact, given that the potential energy is mainly determined by the amplitudes of the fields $s$ and $h$ (if the amplitude of $h$ is non-zero), it can be considered that where the potential takes a larger value in both the $s$ direction ($h$=0) and the $h$ direction ($s=0$) is the region where the domain wall exists. Specifically, if the sign of $s$ differs at two neighboring grid points separated by $\delta x$, it indicates that the amplitude of $s$ is zero at a certain position between the two grid points, then the potential energy takes a local maximum value in the $s$ direction here. At the same time, if the absolute value of the amplitude of $h$ is small on these two adjacent lattice points, for example, if $h<0.2v$ is satisfied, then the value of the potential energy in the $h$ direction is large here. In summary, if the absolute value of $h$ is less than $0.2v$ as well as the sign of $s$ is different on two adjacent lattice points, the domain walls intersect the link between the two points. 

The scaling parameter $\xi_{\rm dw}$ (or equivalently the area parameter $\mathcal{A}$) of DWs is defined as 
\begin{equation}
\xi_{\rm dw}\equiv\mathcal{A}=\frac{\rho_{\rm wall}}{\sigma_{\rm wall}}t, \ \ \text{with} \ \rho_{\rm wall}=\frac{\sigma_{\rm wall}A}{a(t)V},
\end{equation}
where $\rho_{\rm wall}$ is the energy density of the domain wall, $\sigma_{\rm wall}$ is the surface mass density of the domain wall, $A$ and $V$ are the comoving area of domain wall and the comoving volume of the simulation box, respectively. The simplified scaling parameter can be written as
\begin{equation}
\xi_{\rm dw}=\frac{At}{a(t)V}.
\end{equation}
To calculate the comoving area of the domain wall, we use the algorithm proposed by Press, Ryden, and
Spergel~\cite{Press:1989yh}. 
We define the quantity $\delta$ that takes the value 1 at both ends of the link intersecting with the domain wall and equals 0 elsewhere. Then, the comoving area can be expressed as 
\begin{equation}
A=\Delta A \sum\limits_{\rm links} \delta \frac{|\nabla s|}{|s_{,x}|+|s_{,y}|+|s_{,z}|},
\end{equation}
where $\Delta A=(\delta x)^2$ is the comoving area of one grid surface, and $s_{,i}(i=x,y,z)$ are the spatial derivatives of $s(\boldsymbol{{\rm x}})$.

\section{Calculation of GW energy density power spectrum}
Gravitational waves are usually described by spatial metric perturbations $h_{ij}$ around the FLRW background metric
\begin{equation}
{\rm d}s^2=-{\rm d}t^2+a^2(t)(\delta_{ij}+h_{ij}){\rm d}x^i{\rm d}x^j,
\end{equation}
where $t$ represents cosmic time, $h_{ij}$ are transverse and traceless which satisfying $\partial_i h_{ij}=0$ and $h_{ii}=0$.
The equations of motion for $h_{ij}$ is 
\begin{equation}
\ddot{h}_{ij}+3\frac{\dot{a}}{a}\dot{h}_{ij}-\frac{\nabla^2}{a^2}h_{ij}=\frac{16\pi G}{a^2}\Pi_{ij}^{\rm TT},
\end{equation}
where $\dot{} = {\rm d}/{\rm d}t$ and $\nabla_i={\rm d}/{\rm d}x^i$, $G$ is the Newton's constant, and $\Pi_{ij}^{\rm TT}$ is the transverse-traceless(TT) part of the anisotropic energy-momentum tensor $\Pi_{ij}$. The anisotropic tensor $\Pi_{ij}$ can be characterized by the deviation of the field energy-momentum tensor $T_{\mu \nu}$ from perfect fluid
\begin{align}
\Pi_{ij} &\equiv T_{ij}-pg_{ij}=T_{ij}-\frac{\delta_{ij}}{3} \sum_{l}T_{ll} \\ 
&= (\partial_i s \partial_j s+\partial_i h \partial_j h) -\frac{\delta_{ij}}{3} \sum_{l}(\partial_l s \partial_l s+\partial_l h \partial_l h),
\end{align}
where $p$ is the homogeneous background pressure and $g_{ij}=a^2(\delta_{ij}+h_{ij})$. The last term in Eq.(S23) vanishes after transverse-traceless projection. In fact, it is difficult to do a transverse-traceless projection for a tensor in configuration space, given that this projection is a non-local operation. It is more convenient to handle it in Fourier space, and the projection operation can be expressed as 
\begin{equation}
\Pi_{ij}^{\rm TT}=\Lambda_{ij,kl}(\hat{\boldsymbol{{\rm k}}})\Pi_{kl}(\hat{\boldsymbol{{\rm k}}},t).
\end{equation}
The projection operator $\Lambda_{ij,kl}(\hat{\boldsymbol{{\rm k}}})$ is defined as
\begin{equation}
\Lambda_{ij,lm}(\hat{\boldsymbol{{\rm k}}}) = P_{il}(\hat{\boldsymbol{{\rm k}}}) P_{jm}(\hat{\boldsymbol{{\rm k}}}) - \frac{1}{2}P_{ij}(\hat{\boldsymbol{{\rm k}}}) P_{lm}(\hat{\boldsymbol{{\rm k}}}), \quad \text{with} \ P_{ij}=\delta_{ij}- \hat{\boldsymbol{{\rm k}}}_{i} \hat{\boldsymbol{{\rm k}}}_{j},\quad \hat{\boldsymbol{{\rm k}}}_{i}=\boldsymbol{{\rm k}}_{i}/k.
\end{equation}
It can be verified that the transverse-traceless conditions, $k_i \Pi_{ij}(\hat{\boldsymbol{{\rm k}}},t) =\Pi_{ii}(\hat{\boldsymbol{{\rm k}}},t)=0$ are satisfied at all times in momentum space.

To calculate the GW energy density power spectrum, we need to obtain the TT tensor perturbations in momentum space, namely $h_{ij}(k,t)$. A scheme convenient for numerical simulation is proposed to compute $h_{ij}(k,t)$. That scheme introduces unphysical fields $u_{ij}(\boldsymbol{{\rm x}},t)$, which is the solution of the following equation
\begin{equation}
\ddot{u}_{ij}+3H\dot{u}_{ij}-\frac{\nabla^2}{a^2}u_{ij}=\frac{16\pi G}{a^2}\Pi_{ij}.
\end{equation}
We can evolve $u_{ij}(\boldsymbol{{\rm x}},t)$ in configuration space until the time when we need to calculate the GW power spectrum with $h_{ij}(k,t)$. Then we apply Fourier transform to  $u_{ij}(\boldsymbol{{\rm x}},t)$ to obtain $u_{ij}(\boldsymbol{{\rm k}},t)$ in momentum space, and finally we can obtain $h_{ij}(k,t)$ through TT projection
\begin{equation}
h_{ij}(k,t)=\Lambda_{ij,kl}(k)u_{kl}(k,t).
\end{equation}
By implementing the above scheme, we can achieve the same results as the Fourier transform of the solution of Eq.(S21), while avoiding doing Fourier transform at each time step of evolution.

The energy density of a stochastic GW background (SGWB) is usually defined as the 00 component of the stress-energy tensor $t_{\mu \nu}$:
\begin{align}
\rho_{\rm GW}(t) =t_{00}= \frac{1}{32\pi G}\langle \partial_{\mu}{h}_{ij}(\boldsymbol{{\rm x}},t) \partial_{\nu}{h}^{ij}(\boldsymbol{{\rm x}},t) \rangle|_{\mu=\nu=0} = \frac{1}{32\pi G}\langle \dot{h}_{ij}(\boldsymbol{{\rm x}},t) \dot{h}^{ij}(\boldsymbol{{\rm x}},t) \rangle_{V} , 
\end{align}
where the bracket $\langle \cdots \rangle_{V}$ denotes spatial average over a volume $V$.  In momentum space, when $kV^{\frac{1}{3}} \gg 1$ is satisfied, the GW energy density in Eq.(S28) can be further expressed as
\begin{align}
\rho_{\rm GW}(t) &= \frac{1}{32\pi GV} \int_{V}{\frac{{\rm d}^3 \boldsymbol{{\rm k}}}{(2\pi)^3} \dot{h}_{ij}(\boldsymbol{{\rm k}},t) \dot{h}_{ij}^*(\boldsymbol{{\rm k}},t) } \\  &=  \frac{1}{32\pi GV}\frac{1}{(2\pi)^3} \int{ |\boldsymbol{{\rm k}}|^3{\rm d(ln}k)} \int{ {\rm d} \Omega |\dot{h}_{ij}(\boldsymbol{{\rm k}},t)|^2}   \\& \equiv \int{ \frac{{\rm d} \rho_{\rm GW}}{{\rm d}{\rm ln}k} {\rm d}{\rm ln}k} , 
\end{align}
where ${\rm d}\Omega$ represents a solid angle measure in momentum space. 
The GW energy density per logarithmic interval can be obtained from Eq.(S30)
\begin{align}
\frac{{\rm d} \rho_{\rm GW}}{{\rm d}{\rm ln}k} = \frac{k^3}{(4\pi)^3GV}\int{ \frac{{\rm d}\Omega}{4\pi} |\dot{h}_{ij}(\boldsymbol{{\rm k}},t)|^2}.
\end{align}

If gravitational waves are radiated by stochastic sources, the spatial average is replaced by an ensemble average $\langle \cdots \rangle$ over realizations of the stochastic background
\begin{align}
\rho_{\rm GW}(t) &= \frac{1}{32\pi G}\langle \dot{h}_{ij}(\boldsymbol{{\rm x}},t) \dot{h}^{*}_{ij}(\boldsymbol{{\rm x}},t) \rangle \\ &= \frac{1}{32\pi G} \int{\frac{{\rm d}^3 \boldsymbol{{\rm k}}}{(2\pi)^3} \frac{{\rm d}^3 \boldsymbol{{\rm k}'}}{(2\pi)^3} e^{-i\boldsymbol{{\rm x}}\cdot (\boldsymbol{{\rm k}}-\boldsymbol{{\rm k}}') } \times \langle \dot{h}_{ij}(\boldsymbol{{\rm k}},t) \dot{h}_{ij}(\boldsymbol{{\rm k'}},t) \rangle} \\  &=  \frac{1}{(4\pi)^3G} \int{\frac{{\rm d}k}{k}k^3P_{\dot{h}}(k,t)}, 
\end{align}
where the spectrum of the tensor time derivative is defined as $\langle \dot{h}_{ij}(\boldsymbol{{\rm k}},t) \dot{h}_{ij}(\boldsymbol{{\rm k'}},t) \rangle=(2\pi)^3P_{\dot{h}}(k,t)\delta^{(3)}(\boldsymbol{{\rm k}}-\boldsymbol{{\rm k'}})$ under the premise of homogeneity and isotropy. 

The GW energy density power spectrum is defined as energy density per logarithmic interval and is typically normalized by the critical energy density, $\rho_{c}\equiv3H^2/8\pi G$. With Eq.(S35), the dimensionless GW power spectrum is finally expressed as
\begin{align} 
\Omega_{\rm GW} \equiv \frac{1}{\rho_{c}} \frac{{\rm d}\rho_{\rm GW}}{{\rm dln}k} = \frac{1}{\rho_{c}} \frac{k^3}{(4\pi)^3G}P_{\dot{h}}(k,t).
\end{align}

After getting the GW power spectrum at the end time of our simulation, we can convert it into today's spectrum. We follow the scheme described in Ref.~\cite{Dufaux:2007pt,Easther:2007vj,Price:2008hq}.  Firstly, due to the increase in spatial volume and the decrease in GW frequency, the GW energy density scales as
\begin{align} 
\frac{\rho_{\rm GW,0}}{\rho_{\rm GW,e}}=\frac{a_{\rm e}^4}{a_{0}^4}.
\end{align}
The subscripts "0" and "e" denote the quantities today and at the end of our simulation, respectively, and the same is true for the following parts. We assume that entropy is always conserved, so the radiation energy density satisfies
\begin{align} 
\frac{\rho_{\rm rad,0}}{\rho_{\rm rad,e}}=\frac{a_{\rm e}^4}{a_{0}^4}\frac{g_{\rm e}^{1/3}}{g_{0}^{1/3}},
\end{align}
with $g_{\rm e} (g_{0})$ the number of relativistic degrees of freedom at the end time of our simulation (today). We take $g_{\rm e}=106.5$ at the end of our simulation, and $g_{0}=3.36$ today. In the radiation-dominated era, the radiation energy density at the end of our simulation is approximately equal to the critical energy density, that is, $\rho_{\rm rad,e}\approx \rho_{c,\rm e}=3H_{\rm e}^2/8\pi G$, where $H_{\rm e}$ is the Hubble parameter at the end of our simulation. Next, we can obtain the physical frequency of gravitational waves today, which is
\begin{align} 
f_{0}&=\frac{k_{\rm p,0}}{2\pi}=\frac{k_{\rm co,e}}{2\pi a_{\rm e}}\frac{a_{\rm e}}{a_{0}}=\frac{k_{\rm co,e}}{2\pi a_{\rm e}}\left(\frac{\rho_{\rm rad,0}}{\rho_{\rm rad,e}}\right)^{1/4}\left(\frac{g_0}{g_{\rm e}}\right)^{1/12} \nonumber \\ &\approx \frac{k_{\rm co,e}}{a_{\rm e}\rho_{c,\rm e}^{1/4}}(4.4\times10^{10} {\rm Hz})
\end{align}
where $k_{\rm p,0}$ denote physical momentum today and $k_{\rm co,e}$ denote comoving momentum at the end of our simulation. 
The GW power spectrum today can be expressed as 
\begin{align} 
\Omega_{\rm GW,0}h^2 &= \frac{h^2}{\rho_{c,\rm 0}} \frac{{\rm d}\rho_{\rm GW,0}}{{\rm dln}k_{\rm p,0}} \nonumber \\ &= \Omega_{\rm rad,0}h^2\left(\frac{g_{\rm 0}}{g_{\rm e}}\right)^{1/3}\frac{1}{\rho_{c,\rm e}}\frac{{\rm d}\rho_{\rm GW,e}}{{\rm dln}k_{\rm co,e}},
\end{align}
where $\Omega_{\rm rad,0}h^2=h^2\rho_{\rm rad,0}/\rho_{c,\rm 0}\approx4.3\times10^{-5}$ is the abundance of radiation today. Note that we multiplied by an additional factor $h^{2}$, to take into account the uncertainty of the Hubble expansion rate today. 

\section{Numerical simulation results} 
\label{sec:numerical simulation results}

\begin{figure*}[!htp]
\includegraphics[width=.3\textwidth]{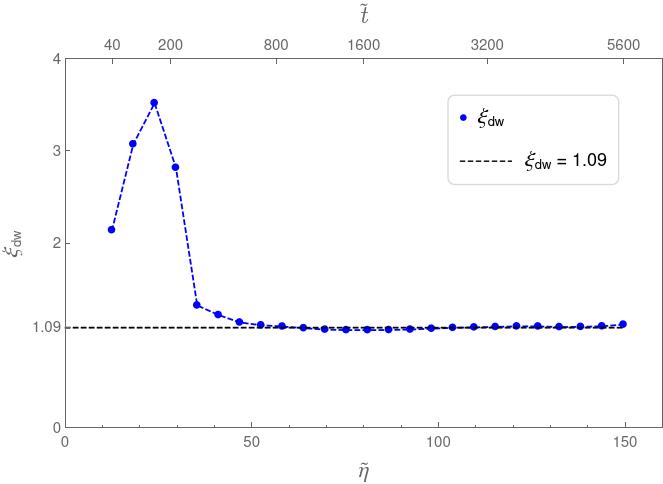}
\hspace{3mm}
\includegraphics[width=.31\textwidth]{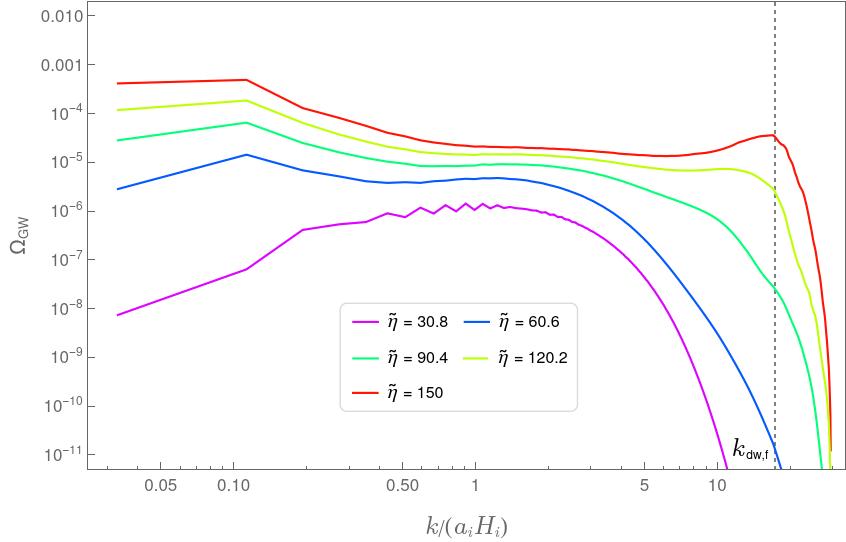} 
\hspace{3mm}
\includegraphics[width=.29\textwidth]{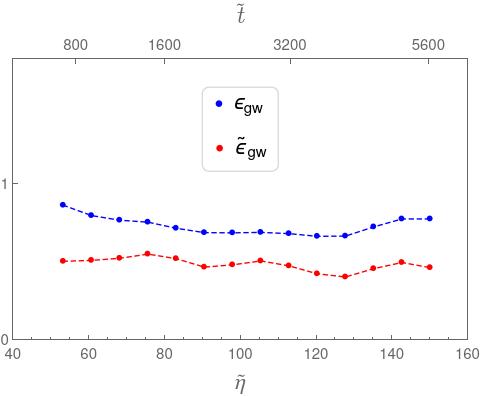} 
\hspace{16mm}
  \caption{Left:Evolution of scaling parameters in rescaled conformal time $\tilde{\eta}$ and dimensionless cosmic time $\tilde{t}$ ($\tilde{t}=w_*/(2H)$) after second order phase transition. The blue solid points are the scaling parameter calculated according to Eq.(S18), and the black dashed line is the result of averaging all scaling parameters after $\tilde{\eta}=50$; Middle: Energy density power spectrum of gravitational waves radiated by pure DW networks at different times. Right: Evolution of the efficiency parameter ($\epsilon_{\rm gw}$) and differential amplitude ($\tilde{\epsilon}_{\rm gw}$) for GWs radiated by DW networks.  }
  \vspace{0.1cm}
  \label{fig:GWspdw}
\end{figure*}

Our numerical simulation results are shown below. For convenience, we denote the number of bubbles of the field $s$ (or field $h$) as $N_{\rm b}$. We simulated three cases in the simulation box with grid points number of $1024^3$, namely pure domain walls (without bubble generation, $N_{\rm b}=0$), and the case where 512 or 64 bubbles ($N_{\rm b}=512$ or $N_{\rm b}=64$) are generated during first-order phase transitions. At first, for the simulation of pure domain walls (without bubble generation and ignoring the field $h$ all the time), we measured the scaling parameters ($\xi_{\rm dw}$) of the domain wall network at different times (see the left panel of Fig.~\ref{fig:GWspdw}) and found that the domain wall network entered the scaling regime after $\tilde{\eta}=50$, and the scaling parameters remained stable at about 1.09. We measured the power spectrum of gravitational waves radiated by pure domain walls (see the middle panel of Fig.~\ref{fig:GWspdw}). The vertical black dashed line in the figure represents the dimensionless comoving wavenumber (momentum) which corresponds to the thickness of the domain wall at the final time
\begin{equation}
k_{\rm dw,f}=(2\pi/\delta_{\rm dw,f}) \times a(\tilde{\eta}_{f}) /(a_iH_i),
\end{equation}
where $\delta_{\rm dw,f}$ is the physical thickness of the domain wall at the final time $\tilde{\eta}_{f}$. The physical thickness of the domain wall can be expressed as $\delta_{\rm dw}=1/m_{\rm s,eff}$, with $m_{\rm s,eff}^2=\partial^2 V(s,0,T)/\partial s^2$ the square of the effective mass of the field $s$. We also measured the efficiency parameter $\epsilon_{\rm gw}$ and differential amplitude $\tilde{\epsilon}_{\rm gw}$ of GWs radiated by the domain-wall network after entering the scaling regime (see the right panel of Fig.~\ref{fig:GWspdw}), they can be expressed as~\cite{Hiramatsu:2013qaa}
\begin{equation}
\epsilon_{\rm gw}\equiv \frac{\rho_{\rm gw}}{G\xi_{\rm dw}^2\sigma_{\rm wall}^2}, \quad \tilde{\epsilon}_{\rm gw}\equiv \left( \frac{{\rm d}\epsilon_{\rm gw}}{{\rm dln}k} \right)_{\rm peak}=\frac{1}{G\xi_{\rm dw}^2\sigma_{\rm wall}^2} \left(\frac{{\rm d}\rho_{\rm gw}(t)}{{\rm dln}k}\right)_{\rm peak},
\end{equation}
where the subscript “peak” implies that the value is computed at the peak of the power spectrum of GWs. We obtain that the average values of $\epsilon_{\rm gw}$ and $\tilde{\epsilon}_{\rm gw}$ are 0.73 and 0.49, respectively.

\begin{figure*}[htb]
\includegraphics[width=.2\textwidth]{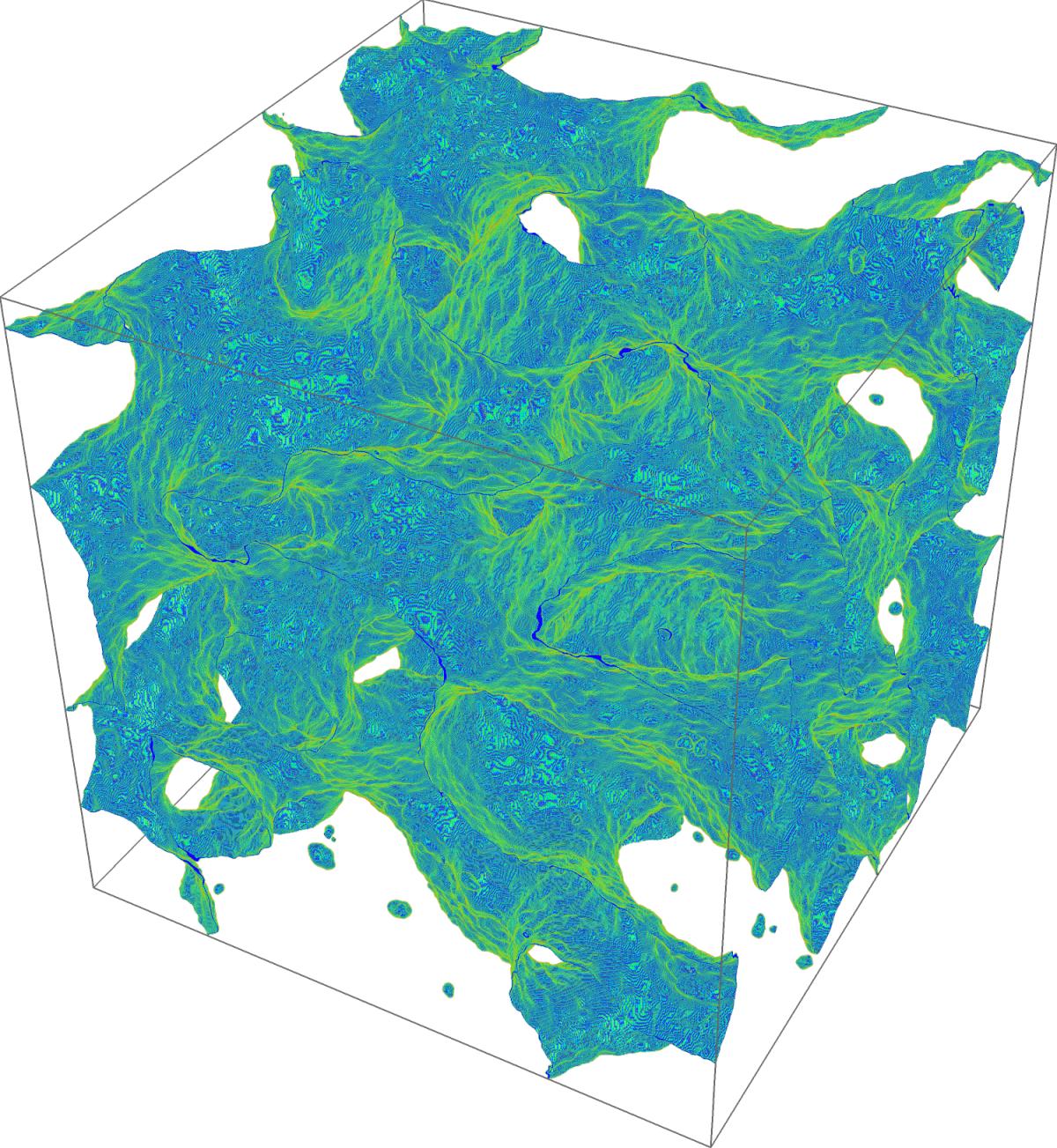} 
\hspace{3mm}
\includegraphics[width=.2\textwidth]{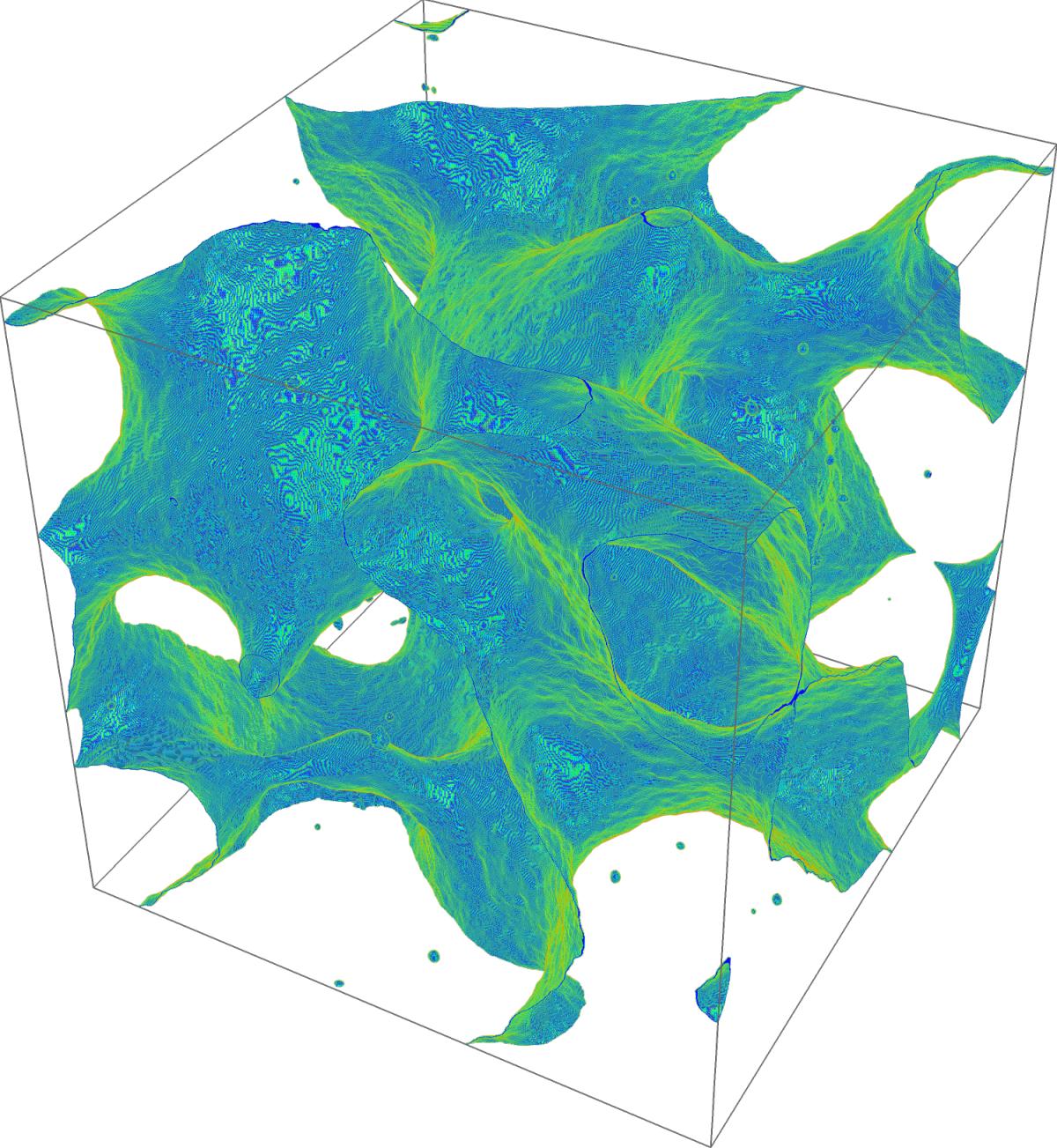}
\hspace{3mm}
\includegraphics[width=.2\textwidth]{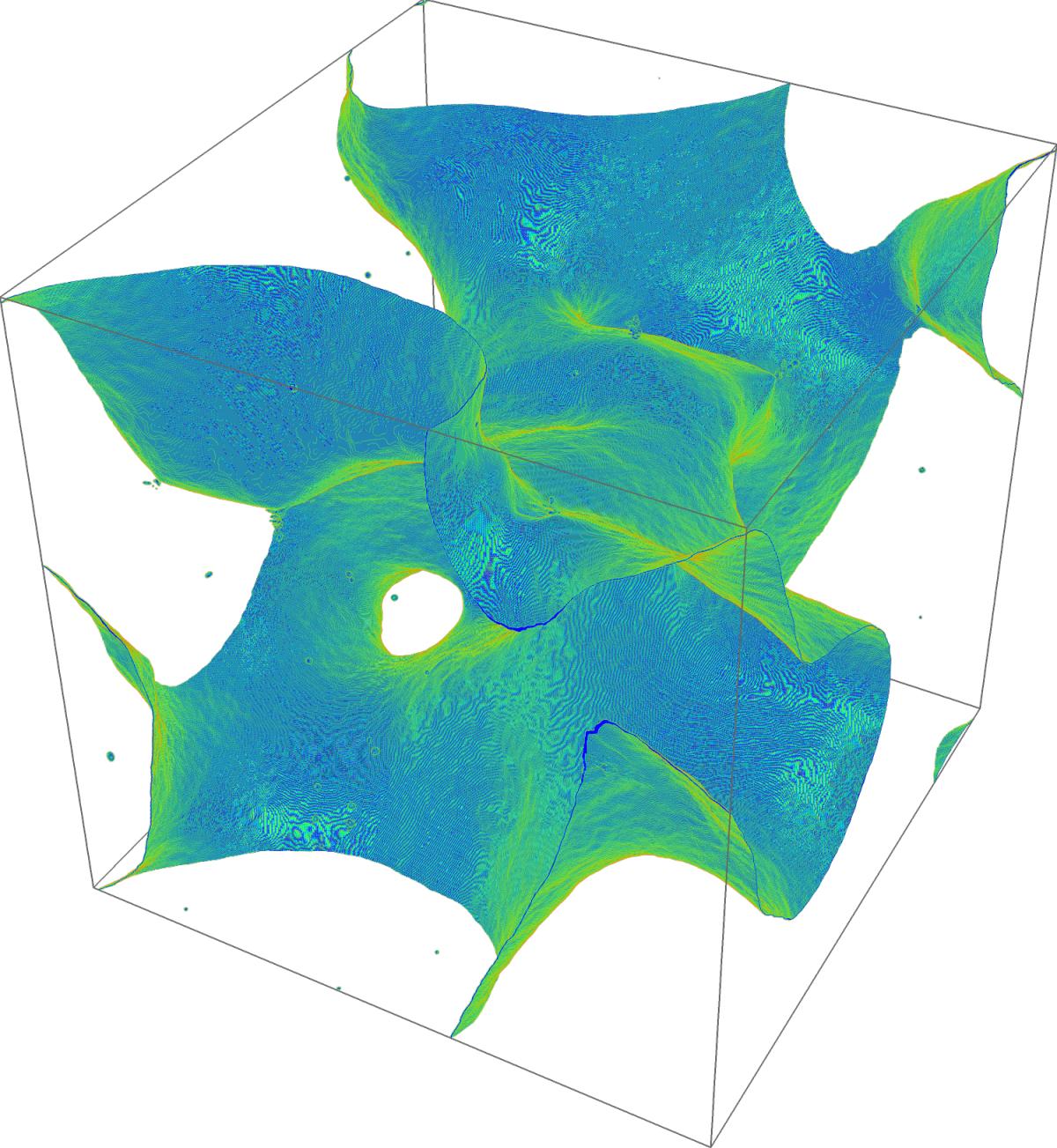}
\hspace{3mm}
\includegraphics[width=.2\textwidth]{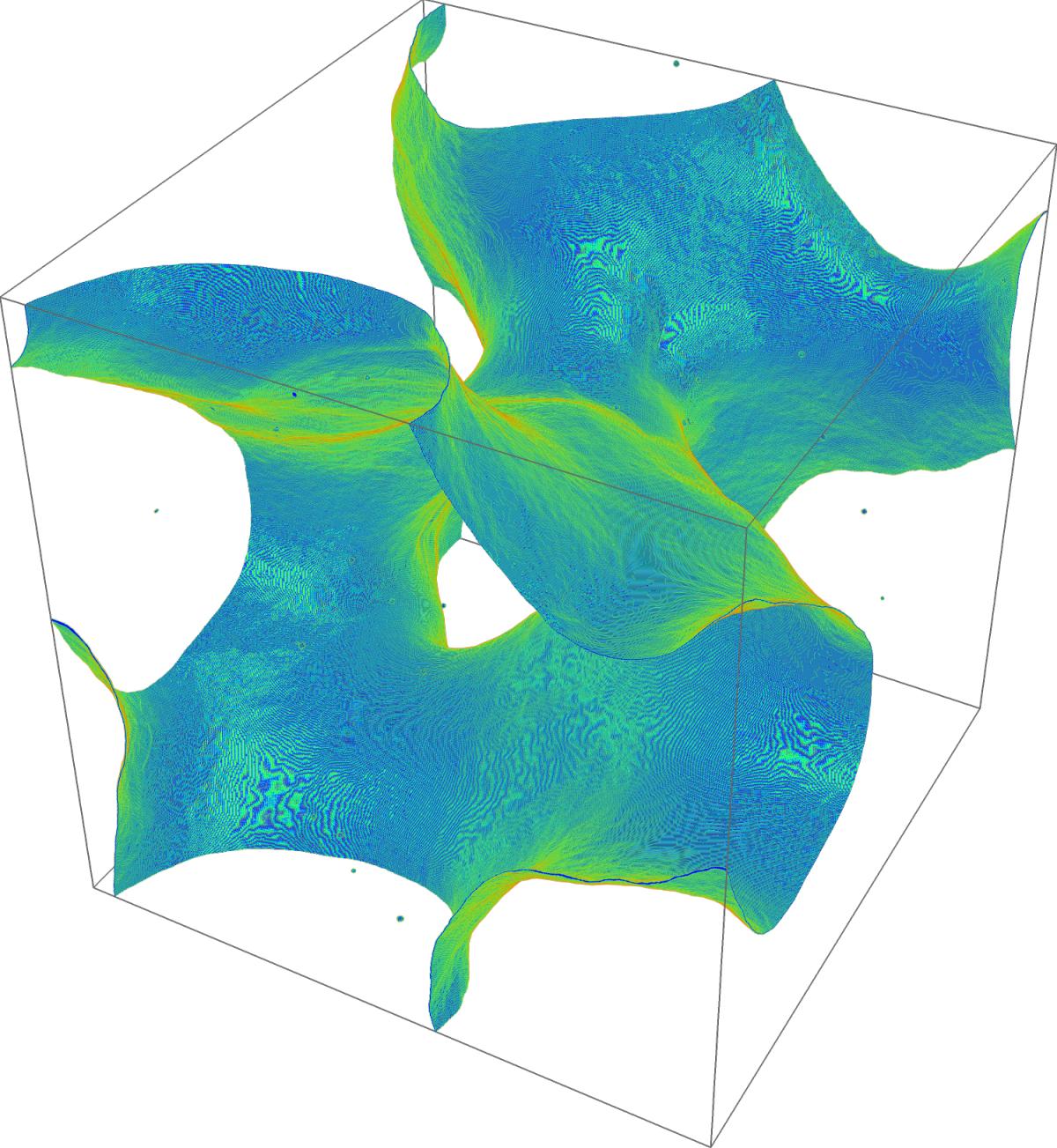}
  \caption{3D snapshots of the pure DW networks at different times, the four plots arranged from left to right were recorded at the time  $\tilde{\eta}$=52.39, 80.94, 120.91, and 149.46, respectively. The blue region indicates where the domain wall exists.}
  \vspace{0.1cm}
  \label{fig:3D snapshots of pure domain wall}
\end{figure*}

For intuitiveness, we recorded the three-dimensional distribution of the pure domain wall networks at different times (see Fig.~\ref{fig:3D snapshots of pure domain wall}).

\begin{figure*}[!htp]
\includegraphics[width=.38\textwidth]{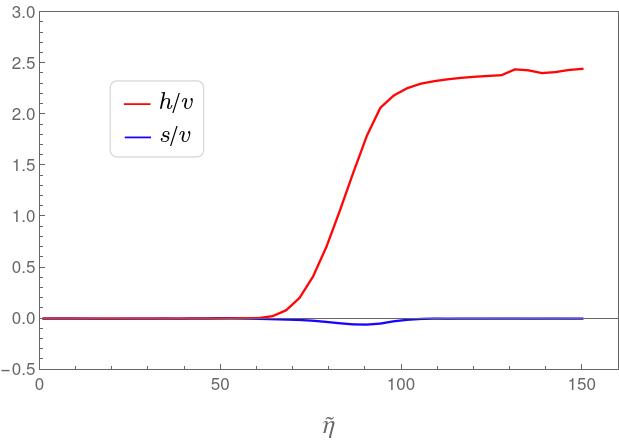}
\hspace{5mm}
\includegraphics[width=.41\textwidth]{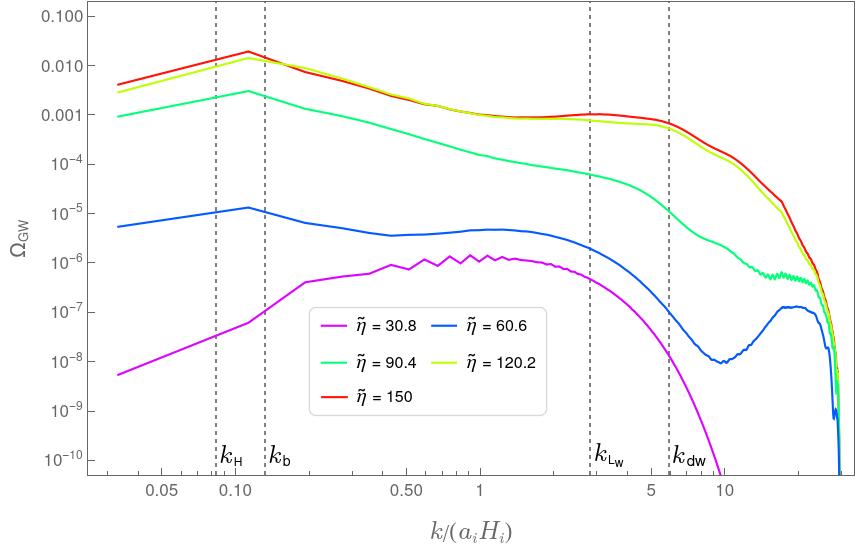} 
\hspace{16mm}
  \caption{Left:  Evolutions in average amplitude of field $s$ and $h$ in the case of $N=1024$ and $N_{\rm b}=64$; Right: GW power spectrum at different times in a two-step phase transition scenario with $N=1024$ and $N_{\rm b}=64$. }
  \vspace{0.1cm}
  \label{fig:gwnb64}
\end{figure*}

For the case where 64 bubbles ($N_{\rm b}=64$) are generated during the first order phase transition, the average amplitude of the fields $s$ and $h$ changes as shown in the left panel of Fig.~\ref{fig:gwnb64}. We also measured the power spectrum of gravitational waves in the right plot.

\begin{figure*}[htb]
\includegraphics[width=.4\textwidth]{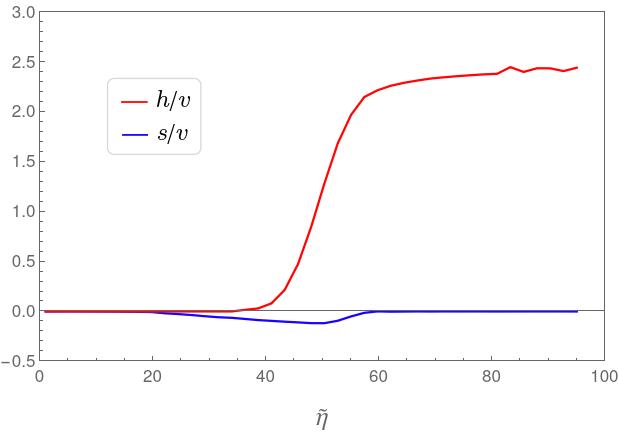} 
\hspace{5mm}
\includegraphics[width=.425\textwidth]{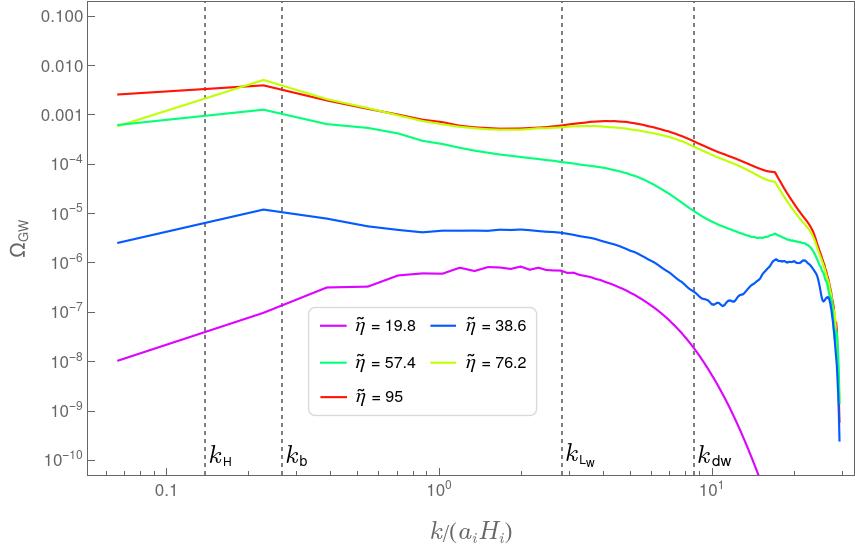} 
\hspace{16mm}
  \caption{ Left: Evolutions in average amplitude of field $s$ and $h$ in the case of $N=512$ and $N_{\rm b}=64$; Right: GW power spectrum measured at different times for the case.   }
  \vspace{0.1cm}
  \label{fig:shgw64}
\end{figure*}

To investigate the impact of the number of grid points ($N$) on our simulation, we performed another set of simulations with $N=512$ points per side. Similarly, We fix $\tilde{\eta}=1$ to be the time at which $T=8T_{\rm c}$, so the second-order phase transition happens at $\tilde{\eta}=8$. We still set $f_*=\eta=6\times10^{16}{\rm GeV}$ and $w_*=a_iH_i$ to do the rescale for dimensional physical quantities.
We use the second-order leap-frog algorithm to evolve the equations of motion in a simulation box of comoving side-length $L=95/(a_{i}H_{i})$. 
 So, the dimensionless comoving lattice spacing is about $\delta \Tilde{x}=0.19$ (which is equal to the $\delta \Tilde{x}$ in the case of $N=1024$), and the dimensionless time-step is chosen as $\delta \tilde{\eta}=0.005$. As the temperature decreases to about $T=0.5v$, that is, when $\tilde{\eta}=34.68$, the first-order phase transition happens, and bubbles form in the $s$ and $h$ fields due to quantum tunneling. We evolve the equations of motion until the final moment $\tilde{\eta}=95$, at this time, the simulation box contains one Hubble volume, and the thickness of the domain walls is about 1.37 times that of the physical lattice spacing. 
 
\begin{figure*}[htb]
\includegraphics[width=.65\textwidth]{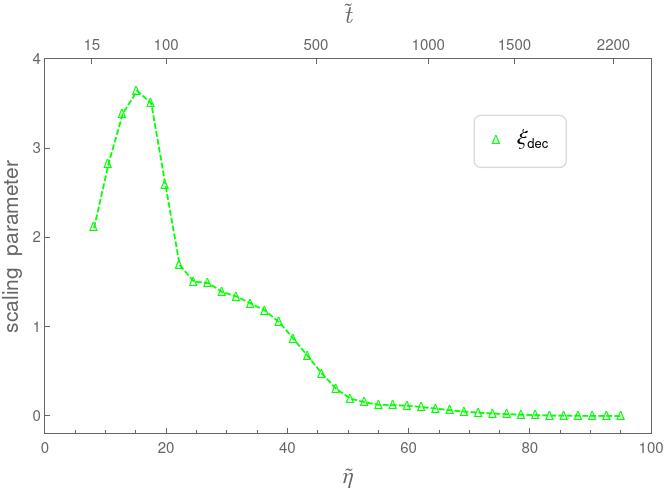} 
\hspace{16mm}
  \caption{Evolution of scaling parameters with rescaled conformal time $\tilde{\eta}$ and dimensionless cosmic time $\tilde{t}$ ($\tilde{t}=w_*/(2H)$) in the case of $N=512$ and $N_{\rm b}=64$.}
  \vspace{0.1cm}
  \label{fig:spnb64}
\end{figure*}

The evolution of the amplitudes of $s$ and $h$ during the simulation is shown in Fig.~\ref{fig:shgw64}.  We also measured the GW energy density power spectrum at five different moments during the simulation process, see the right panel in the Figure.
The evolution of scaling parameters for domain walls is shown in Fig.~\ref{fig:spnb64}.

\end{document}